\documentclass[lettersize,journal]{IEEEtran}
\IEEEoverridecommandlockouts
\usepackage{cite}
\usepackage{amsmath,amssymb,amsfonts}
\usepackage{booktabs}
\usepackage{algorithmic}
\usepackage{graphicx}
\usepackage{enumitem}
\usepackage{textcomp}
\usepackage{hyperref}
\usepackage{float}
\usepackage{soul}
\usepackage{booktabs}
\usepackage{siunitx}
\usepackage{threeparttable}

\usepackage{subcaption}
\usepackage{setspace}
\usepackage{lettrine}
\usepackage{xcolor}
\usepackage[a4paper, total={184mm,239mm}]{geometry}
\def\BibTeX{{\rm B\kern-.05em{\sc i\kern-.025em b}\kern-.08em
    T\kern-.1667em\lower.7ex\hbox{E}\kern-.125emX}}
\begin{document}

\title{YAP+: Pad-Layout-Aware Yield Modeling and Simulation for Hybrid Bonding
}
\renewcommand{\footnoterule}{%
    \hrule width 0.4\textwidth height 0.4pt
    \vspace{0.4em} 
}

\author{\IEEEauthorblockN{Zhichao Chen, Puneet Gupta, \textit{Fellow, IEEE}
}}

\setlength{\baselineskip}{11pt} 
\setlength{\abovedisplayskip}{3pt} 
\setlength{\belowdisplayskip}{3pt} 
\maketitle
\setlength{\parskip}{0pt}
\begin{abstract}
Three-dimensional (3D) integration continues to advance Moore’s Law by facilitating dense interconnects and enabling multi-tier system architectures. Among the various integration approaches, Cu-Cu hybrid bonding has emerged as a leading solution for achieving high interconnect density in chiplet integration. In this work, we present YAP+, a yield modeling framework specifically tailored for wafer-to-wafer (W2W) and die-to-wafer (D2W) hybrid bonding processes.  YAP+ incorporates a comprehensive set of yield-impacting failure mechanisms, including overlay misalignment, particle defects, Cu recess variations, surface roughness, and Cu pad density. Furthermore, YAP+ supports pad layout-aware yield analysis, considering critical, redundant, and dummy pads across arbitrary 2D physical layout patterns. To support practical evaluation, we developed an open-source yield simulator, demonstrating that our near-analytical model matches simulation accuracy while achieving over 1,000x speedup in runtime. This performance makes YAP+ a valuable tool for co-optimizing packaging technologies, assembly design rules, and system-level design strategies. Beyond W2W-D2W comparisons, we leverage YAP+ to investigate the impact of pad layout patterns, bonding pitch, pad ratios across different pad types, and explore the benefits of strategically placing redundant pad replicas.
\end{abstract}

\label{0_abstract}

\section{Introduction}
\lettrine{A}{s} the physical and economic boundaries of scaling traditional two-dimensional integrated circuits are increasingly challenged, Three-Dimensional Integrated Circuits (3D-ICs) have emerged as a compelling alternative to sustain the progression of Moore’s Law. 
By vertically stacking multiple device layers, 3D-ICs offer notable advantages, including shorter interconnect paths, improved performance, reduced power consumption, and higher integration density. 
A key enabler of 3D-IC technology is the hybrid bonding (HB) process. 
Compared to earlier packaging technologies, HB offers a dramatic leap in interconnect density, reaching 10,000 to 1 million connections per \SI{}{\milli\meter\squared}, with sub-micron alignment accuracy down to \SI{50}{\nano\meter} \cite{Mitsuishi2023,Sano2025,Ryan2025}, and ultra-low energy consumption below 0.05 pJ/bit due to the lower resistance of direct Cu connections \cite{Dylan2024}. 
 HB supports fine-pitch, high-reliability interconnects, making it ideal for applications such as high-bandwidth memory, logic-memory integration, and advanced sensing systems \cite{Ikegami2024, Elsherbini2021, Lau2023, Lee2022, Park2022}.
The two predominant HB approaches are wafer-to-wafer (W2W) and die-to-wafer (D2W). 
D2W provides enhanced flexibility by allowing verified top dies to be bonded onto known-good base dies, thereby improving overall yield \cite{Elsherbini2021}. 
In contrast, W2W bonding is more efficient for high-volume production and offers better alignment accuracy, but its yield is more vulnerable to defects in either wafer.
These distinctions in process complexity and yield implications underscore the need for a thorough comparative analysis.

Accurate and predictive yield modeling is critical for advanced packaging technologies, as it enables early identification of potential failure mechanisms based on the design and process factors during the development cycle, facilitates system-technology co-optimization, and informs strategies for chiplet interconnect repair \cite{Marinissen2024}. 
The overall yield in advanced integration schemes is influenced by several components, including the individual yields of chiplets, HB process, and through-silicon vias (TSVs).
While system-level yield modeling has received considerable attention \cite{Campbell2010, Singh2014, Xu2012, Graening2023, Agnesina2023}, existing models for HB yield are often overly simplified.
For instance, \cite{Singh2014, Xu2012} propose yield models tailored to 3D stacked ICs, yet \cite{Singh2014} omits the bonding process entirely, and \cite{Xu2012} treats bonding yield as a fixed constant, which is a simplification also adopted by \cite{Graening2023} and \cite{Agnesina2023} in their chiplet system yield analyses. 
These approaches fail to capture the intricate physical failure mechanisms inherent to HB and do not provide a detailed, process-aware yield model.

In addition to the presence of multiple failure mechanisms, the pad layout pattern, encompassing the ratio of different I/O pad types, their spatial distribution, and the redundancy scheme, can significantly influence the actual yield.
Depending on the pitch and die size, a single die fabricated using the HB process may contain anywhere from hundreds of thousands to several million I/O pads, including signal pads, power/ground pads, and a large number of dummy pads.
The ratio and spatial placement of these types of I/O pads are highly design-dependent.
Signal pads are typically critical, as the failure of even a single signal pad can lead to die failure.
\cite{Zhao2011} introduced a redundant TSV grouping technique to enhance the yield of 3D ICs; similarly, introducing redundancy for signal pads can be an effective strategy to mitigate yield loss in HB, particularly in the presence of large-scale clustering void defects.
In terms of adding redundancy, it can be roughly categorized as \textit{shared redundancy} and \textit{dedicated redundancy}.
A shared redundancy scheme is a fault-tolerance approach where a group of $N$ components share a smaller pool of $M$ spare components \cite{Singh1988, Koren1998}. 
Typically, the value of $M$ is significantly less than $N$ ($M<N$), allowing for a more efficient utilization of spare resources.
A dedicated redundancy scheme is also referred to as 1:1 redundancy, where each main component is paired with its own exclusive, dedicated spare component.
The interaction among the redundancy scheme, the physical distance between main and redundant components, and defect clustering patterns can have various impacts on yield.
Power and ground pads, on the other hand, are often replicated and redundantly distributed across the die to maintain power integrity.
Dummy pads, which share the same pitch and size as functional I/O pads, play a key role in process optimization \cite{Lau2025, Mariappan2024, Kim2022PadStructure}, especially for chemical mechanical planarization (CMP). These pads are strategically inserted to ensure uniform pattern density across the die, with surface coverage typically ranging from 40\% to 90\%, leading to improved surface planarity \cite{US9666566B1}. Dummy pad failures are non-critical to die functionality.
Given that the impact of failures across diverse I/O pad types on die functionality can be highly variable and design-dependent, it is crucial to incorporate pad-layout information into the yield analysis for each specific design.

The complexity of the interaction between multiple failure mechanisms and the pad layout patterns in HB makes analytical modeling of yield challenging.
In response to this challenge, this work extends our previous work \cite{Chen2025}, and proposes an enhanced version, YAP+. YAP+ is a physical mechanism-driven and pad-layout-aware near-analytical yield modeling framework. 
YAP+ introduces a detailed analysis and modeling framework capable of predicting bonding yield for arbitrary pad layouts, which is an important feature absent in YAP.
The code of the yield model (YAP+) and simulator is available open-source at \href{https://github.com/nanocad-lab/YAP}{\texttt{https://github.com/nanocad-lab/YAP}}.
The key contributions of this work are outlined below:
\begin{itemize}
    \item To the best of our knowledge, this is the first yield model specifically designed for the HB process. This model captures key failure mechanisms that contribute to yield loss, including overlay errors, particle defects, Cu recess variations, dielectric surface roughness, and excessive Cu pattern density.
    \item We propose a dilation-based method capable of adaptively calculating the critical area for arbitrary defect shapes and pad layouts that include critical, redundant, and dummy I/O pads. 
    \item We develop a yield simulator based on the statistical distributions of various failure mechanisms to validate the proposed yield model and evaluate its predictive accuracy. 
    \item We conduct detailed case studies to analyze the influence of process and design parameters on yield. These include comparisons between W2W and D2W HB approaches and demonstrate the importance of process control in achieving high yield performance.
\end{itemize}

The remainder of the paper is organized as follows: 
Section II discusses the key failure mechanisms associated with HB processes. 
Section III introduces yield modeling methodologies for W2W HB and extends it to D2W HB. 
Section IV describes the experimental setup, details the Monte Carlo simulation for multiple failure mechanisms, and compares the simulation outcomes with the near-analytical model.
Section V presents case studies analyzing the impact of design and process parameters on yield, including a comparative evaluation of W2W and D2W bonding approaches. 
Section VI concludes the paper and highlights potential directions for future research.

\label{1_introduction}

\section{Overview of Failure Mechanisms of Hybrid Bonding}
This section presents an overview of the key failure mechanisms inherent to HB processes, including overlay misalignment, Cu recess variation, and particle-induced void defects. 
Each of these factors can significantly cause yield loss if the bonding process parameters are not properly controlled. 
A thorough understanding of these mechanisms is essential for optimizing HB process reliability and improving the bonding yield.
\subsection{Overlay Errors}
\begin{figure}[t]
    \centering
    \begin{subfigure}[b]{0.58\linewidth}
        \centering
        \includegraphics[width=\linewidth]{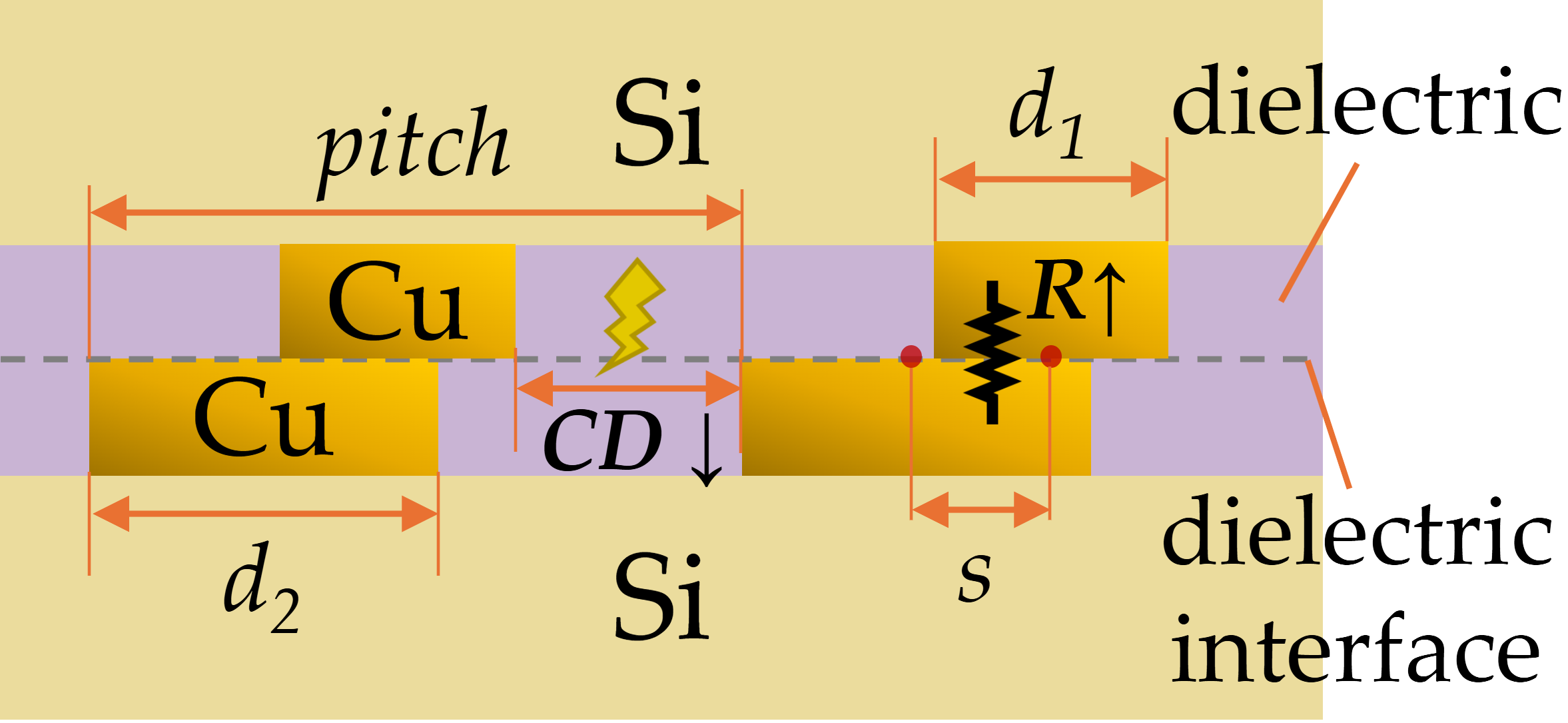}
        \caption{Misalignment leads to resistance increase and dielectric breakdown.}
        \label{fig:overlay_failure_mechanism}
    \end{subfigure}
    \hfill
    \begin{subfigure}[b]{0.4\linewidth}
        \centering
        \includegraphics[width=\linewidth]{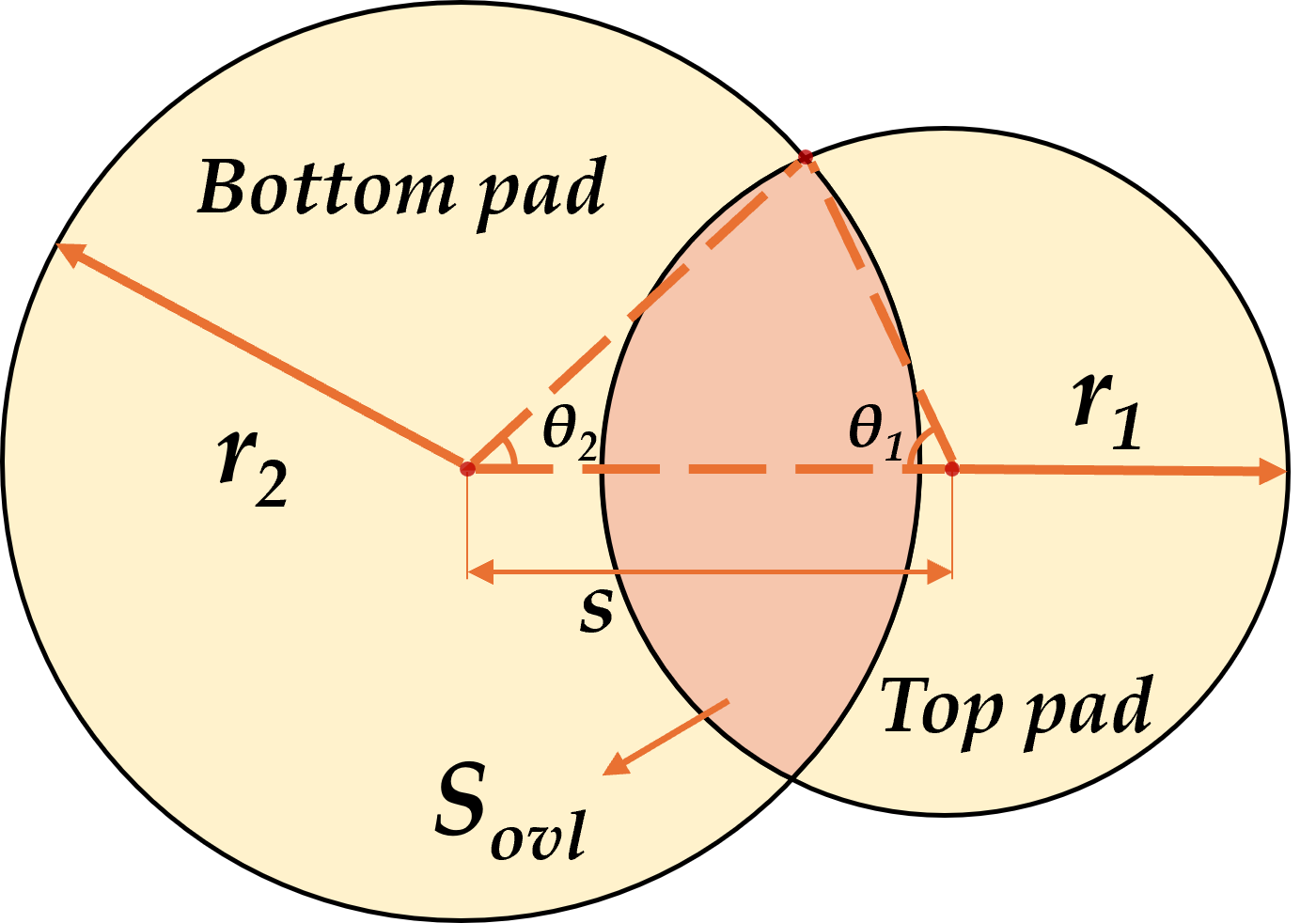}
        \caption{Contact area ($S_{ovl}$) calculation given the misalignment ($s$).}
        \label{fig:overlay_contact_area}
    \end{subfigure}
    \caption{Failure mechanism of Overlay errors.}
    \label{fig:overlay_error}
    \vspace{-10pt}
\end{figure}
Ensuring the quality of the Cu connections formed during the HB process is essential for maintaining the electrical performance of the overall design. 
However, misalignment between Cu pillars of the top and bottom wafers is inevitable due to factors such as robot arm calibration errors and wafer warpage induced by thermal stress mismatches. 
As pad dimensions shrink and bonding pitch achieves the sub-micron scale, the impact of such misalignment on yield becomes increasingly significant \cite{Ryan2025, Zhang2024}.
Fig. \ref{fig:overlay_failure_mechanism} and Fig. \ref{fig:overlay_contact_area} show the front and top views of the bonding connection, respectively.
As shown in the figures, the excessive misalignment ($s$) will decrease the contact area ($S_{ovl}$) of the Cu interface. This reduction leads to increased contact resistance and elevates the risk of electromigration-induced failures \cite{Moreau2023}. 
Additionally, the probability of dielectric breakdown increases as the critical distance ($CD$) between adjacent Cu pads decreases, resulting in a thinner insulating film between the upper pads and the lower pads of neighboring pillars \cite{Ikegami2024}. 
Let the bonding pitch be denoted by $p$, and assume the pads are circular with diameters $d_1=2r_1$ for the top pad and $d_2=2r_2$ for the bottom pad. 
The critical distance between two perfectly aligned Cu pillars is defined as $CD=p-d_2$, representing the spacing between adjacent pads. 
In certain designs, the top pad is intentionally made smaller than the bottom pad to enhance tolerance against misalignment \cite{Kim2020}.
YAP+ supports modeling such asymmetric pad dimension configurations, enabling analysis across a broad spectrum of real-world bonding scenarios.
Overlay errors are generally categorized into pad-level random misalignment and systematic misalignment.
To reduce the likelihood of Cu-to-Cu bonding failure, it is recommended that the total misalignment remain within $50$\% of the bottom pad's diameter \cite{Ryan2025}. 
This constraint becomes increasingly challenging to meet in fine-pitch designs.

\subsection{Cu Recess Variations}
\begin{figure}[t]
    \centering
    \begin{subfigure}[b]{\linewidth}
        \centering
        \includegraphics[width=\linewidth]{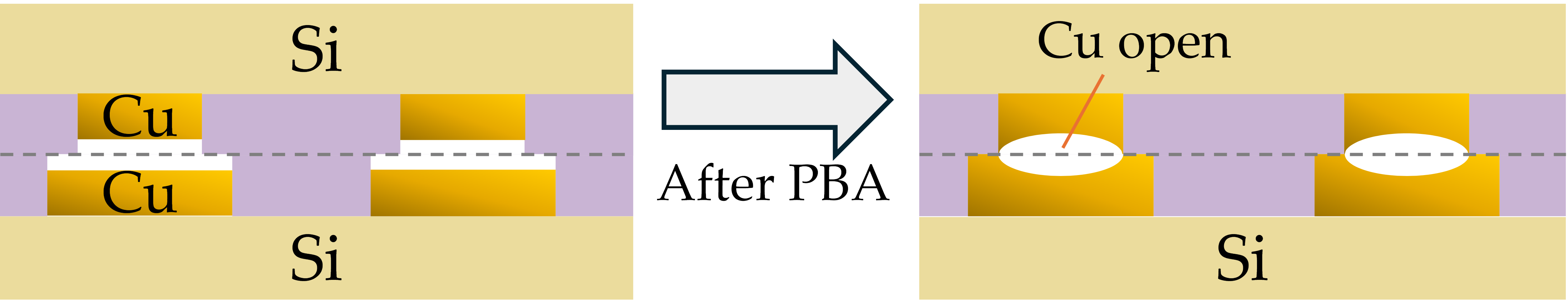}
        \caption{Excessive Cu recess reduces Cu contact area or even cause open.}
        \label{fig:low_Cu_recess}
    \end{subfigure}
    \hfill
    \begin{subfigure}[b]{\linewidth}
        \centering
        \includegraphics[width=\linewidth]{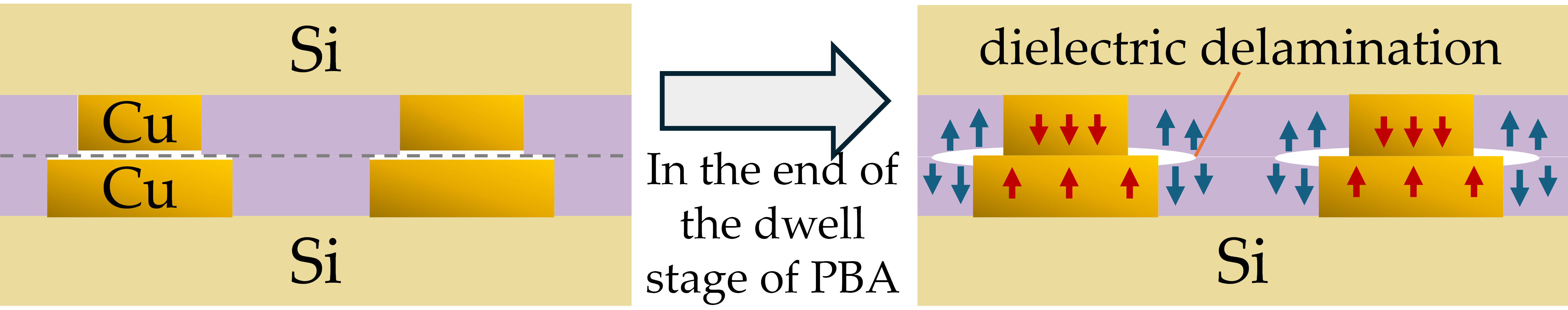}
        \caption{Insufficient Cu recess may cause dielectric delamination, depending on the dielectric bonding strength, Cu density, and post-bond annealing (PBA) temperature, etc. }
        \label{fig:high_Cu_recess}
    \end{subfigure}
    \caption{Failure mechanism of Cu recess variations.}
\end{figure}
The CMP process will introduces Cu recess effects, often resulting in a concave surface profile on the Cu pad. 
As illustrated in Fig. \ref{fig:low_Cu_recess}, excessive Cu recess can degrade the bonding quality or even incur Cu interconnect failure following post-bond annealing (PBA) \cite{Chidambaram2020, Ren2021}. 
Conversely, Cu protrusion and insufficient Cu recess also negatively impact yield.
High wafer surface roughness reduces the effective dielectric contact area during low-temperature bonding, which in turn lowers the density of covalent bonds formed after PBA, weakening both bonding strength and energy per unit area at the dielectric interface \cite{Chidambaram2020, Gui1999, Dubey2024}.
In fine-pitch designs, high Cu pattern density combined with insufficient Cu recess can lead to elevated peak peeling stress at the dielectric interface during the final stage of annealing \cite{Ji2019, Ji2020, Wang2023, Beilliard2017}.
As shown in Fig. \ref{fig:high_Cu_recess}, if the dielectric interface bonding cannot withstand the peeling stress, dielectric delamination or cracking may occur, resulting in bonding failure \cite{Fujii2023, Le2024, Zhao2024}. 
Therefore, to achieve a high yield, especially for chiplets with a large number of Cu pads, a precise control of Cu recess variation across top and bottom pad within a range determined by Cu pattern density, surface roughness, etc. is necessary.

\subsection{Particle Defects}
\begin{figure}[t]
    \centering
    \begin{subfigure}[b]{\linewidth}
        \centering
        \includegraphics[width=\linewidth]{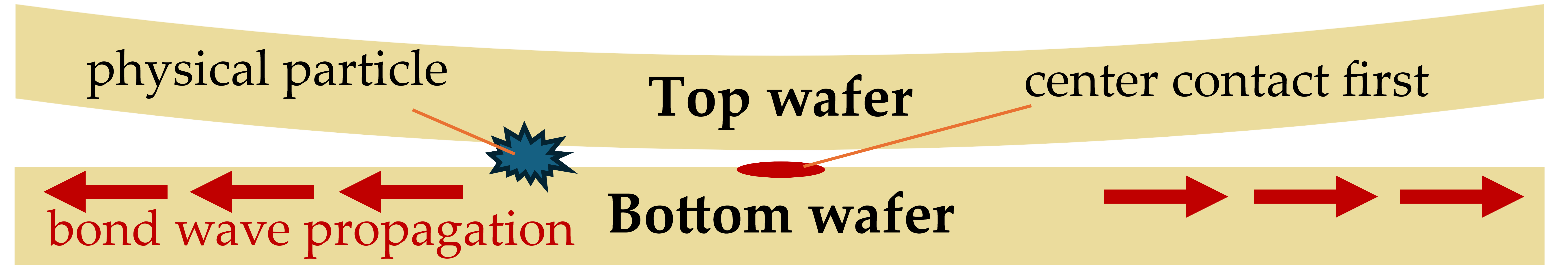}
        \caption{The presence of a particle will form a main void and a void tail due to the bond wave propagation.}
        \label{fig:bond_wave_propagation}
    \end{subfigure}
    \hfill
    \begin{subfigure}[b]{\linewidth}
        \centering
        \includegraphics[width=\linewidth]{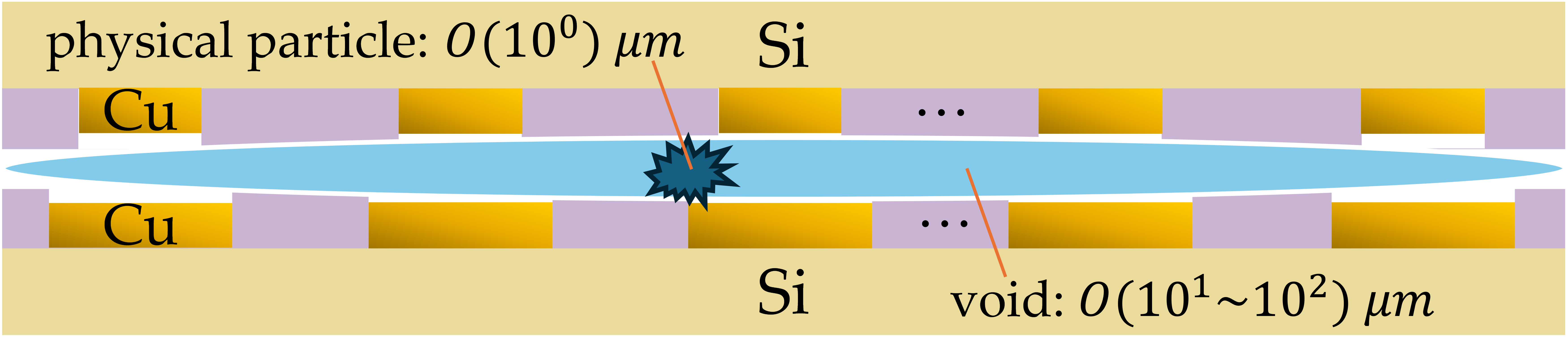}
        \caption{Void formation can fail the dielectric and Cu bonding. A particle of a few microns can form a main void of hundreds of microns \cite{Nagano2022}.}
        \label{fig:particle_void_failure_mechanism}
    \end{subfigure}
    \caption{Failure mechanism of particle defects.}
    \label{fig:particle_failure}
\end{figure}

In the HB process, particles are generated during various steps such as wafer dicing, grinding, and polishing \cite{Xie2025}.
Additionally, any form of friction can produce particles, which is particularly problematic since hybrid bonding involves mechanically picking up dies and placing them onto other chips \cite{Dylan2024}.
Achieving high yield requires stringent cleanliness standards to prevent the presence of physical particles, which can lead to void formation at the bonding interface \cite{Elsherbini2021}. 
Even a particle as small as \SI{1}{\micro\meter} in thickness can cause a void with a diameter reaching hundreds of microns \cite{Nagano2022, Dylan2024}.
In addition to physical particles, gas condensation during the bonding process can incur edge voids near the wafer bevel region. 
However, since the outer edge region is typically removed during the sawing process, dies located away from the wafer perimeter remain unaffected, and thus these voids do not impact overall yield \cite{Kim2022enhancement}.
Consequently, the proposed defect model focuses primarily on yield loss caused by particle-induced void formation.
As shown in Fig. \ref{fig:particle_failure}, during the W2W HB process, initial contact occurs at the center of the top wafer, which then propagates outward toward the edges. 
Due to bond wave propagation, the presence of a particle at the bonding interface can result in a \textit{main void} accompanied by a trailing \textit{void tail} extending radially \cite{Nagano2022}.
In contrast, in the D2W case, void tail formation is uncommon, primarily due to the smaller die size relative to the wafer and differences in the bonding mechanism.

\label{2_overview}

\section{YAP+ Yield Model}
\begin{figure*}[t]
    \centering    \includegraphics[width=\textwidth]{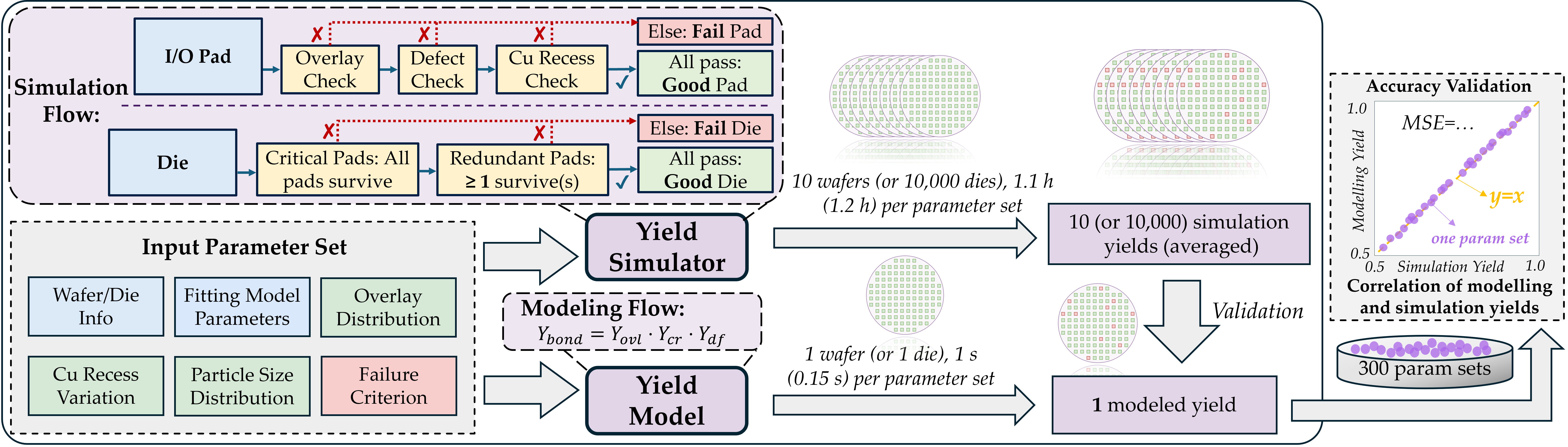}
    \caption{Simulation workflow and the validation of modeling yield on various input parameter sets.}
    \label{fig:simulation_workflow}
    \vspace{-10pt}
\end{figure*}
In this section, we introduce our yield modeling methodology for W2W HB, which incorporates arbitrary pad layout configurations. 
The model is then extended to support D2W bonding scenarios.
Specifically, since power and ground I/O pads typically have numerous replicas distributed widely across the die, it is highly unlikely that a die failure would occur solely due to the simultaneous failure of all such pads.
Therefore, when deriving the yield model below, we do not consider bonding failures of power/ground I/O pads or dummy pads.
To validate our derived model, we compare its predictions against simulation results across 300 distinct parameter sets. 
The simulation and validation workflow is illustrated in Fig. \ref{fig:simulation_workflow}, with detailed experimental settings discussed in Section IV.

\subsection{Overlay Model}
Our proposed overlay model quantifies the yield loss resulting from the misalignment of Cu pads. We assume the bonding misalignment follows a normal distribution with zero mean and a process-dependent $\sigma_1$ \cite{Ghaida2013}. 
Under this assumption, the possibility of survival (POS) of one single pad can be calculated as follows:
\begin{equation}
    POS_{ovl,pad}=\frac{1}{\sigma_1 \sqrt{2 \pi }} \int_{-\delta}^{\delta} e^{-\frac{u^2}{2 \sigma_1^2}}\text{d}u
\end{equation}
where $u$ represents the random overlay error between the top and bottom pads, and $\delta$ denotes the maximum allowable misalignment to ensure the pad's survival. $\delta$ is determined based on the contact area constraint and the critical distance constraint. 
We define $s$ as the systematic overlay error in a Cu connection, which arises from three primary distortion components: \textit{translation}, \textit{rotation}, and \textit{magnification} \cite{Armitage1988}. 
Translation and rotation errors primarily stem from limitations in equipment precision, while magnification errors are mainly caused by wafer warpage/bow due to thermal expansion mismatches among different materials \cite{Ji2019}. 
We define the translation errors in $x,y$ directions as $T_x, T_y$, respectively, and denote the rotation error as $\alpha$. 
Bonded wafer warpage typically ranges from a few micrometers to over \SI{100}{\micro\meter}, but can be reduced to $\sim$\SI{10}{\micro\meter} through run-out compensation techniques \cite{Kang2024}.
Let $B$ denote the warpage of the bonded wafer. 
Studies have shown that the magnification factor $E$ is linearly correlated with $B$ \cite{Kang2024, Okudur2024}.
Based on this observation, we construct a linear model to characterize $E$ as follows:
\begin{equation}
    E=k_{mag}\cdot B
    \label{eq:bow_model}
\end{equation}
where $k_{mag}$ serves as a fitting parameter in the model and is influenced by factors such as the Cu pad depth, Cu pattern density, and the bonding process temperature, etc. \cite{Ji2019}. 
We model the systematic misalignment $\Delta x, \Delta y$ in x, y directions respectively by 
\begin{equation}
    \begin{cases}
    \Delta x(x,y) = T_x - \alpha\cdot y + E\cdot x, \\
    \Delta y(x,y) = T_y + \alpha\cdot x + E\cdot y.  
    \end{cases}
    \label{eq:sys_ovl_xy}
\end{equation}
The systematic overlay error $s$ at the location $(x,y)$ is by
\begin{equation}
    s(x,y)=\sqrt{\left[\Delta x(x,y)\right]^2 + \left[\Delta y(x,y)\right]^2}
    \label{eq:sys_ovl}
\end{equation}
As Fig. \ref{fig:overlay_contact_area} shows, the contact area of two Cu pads can be calculated by
\begin{equation}
\label{eq:ovlp_area}
    S_{ovl}=
    \left\{
    \begin{aligned}
        &\pi r_1^2, &s<r_2-r_1\\
        &\theta_1r_1^2+\theta_2r_2^2-sr_1\sin{\theta_1},& r_2-r_1
        \leq s\leq r_1+r_2.\\
        &0, & s>r_1+r_2
    \end{aligned}
    \right.
\end{equation}
Assuming that pad survival requires the contact area to exceed a threshold defined by $k_{ca}$ times the surface area of the top pad interface, i.e., $S_{ovl}>k_{ca}\pi r_1^2$, and that the critical distance $CD$ must be greater than $k_{cd}$ times the ideal critical distance, i.e., $CD>k_{cd}(p-d_2)$, then $\delta$ can be expressed as
\begin{equation}
    \begin{aligned}
    \delta=&\min\bigg\{\frac{\theta_1r_1^2+\theta_2r_2^2-k_{ca}\pi r_1^2}{r_1\sin\theta_1}, \\
    &(1-k_{cd})p-\frac{1}{2}d_1+\left(k_{cd}-\frac{1}{2}\right)d_2  \bigg\}
    \end{aligned}
    \label{eq:circ_delta}
\end{equation}
In practice, the failure region associated with overlay errors usually spans a distance larger than that separating a redundant pad from its replica.
Consequently, if misalignment causes one redundant pad to fail, its replica is very likely to fail as well.
Therefore, the overall POS for a die is determined by the lowest POS among all interconnection pads, excluding dummy pads and power/ground I/O pads that have numerous, widely distributed replicas.
Given above, say that a die has $N_{cr}$ critical and $N_{rd}$ redundant pads, its POS can be written as
\begin{equation}
    POS_{ovl,die}=\frac{1}{\sigma_1 \sqrt{2 \pi }} \min_{i\in[1,N_{cr}+N_{rd}] }\left\{\int_{-\delta-s_i}^{\delta-s_i} e^{-\frac{u^2}{2 \sigma_1^2}}\text{d}u\right\}
    \label{eq:pos_ovl}
\end{equation}
where $s_i$ denotes the systematic overlay misalignment of the $i$-th interconnection pad (critical or redundant) on the die.
Assuming one wafer has $M$ dies, the overlay yield is by
\begin{equation}
    Y_{ovl,W2W}=\frac{1}{M}\sum^M_{j=1}POS_{ovl,die,j}
    \label{eq:yield_ovl_W2W}
\end{equation}
\begin{figure}[b]
    \centering
    \begin{subfigure}[b]{0.49\linewidth}
        \centering
        \includegraphics[width=\linewidth]{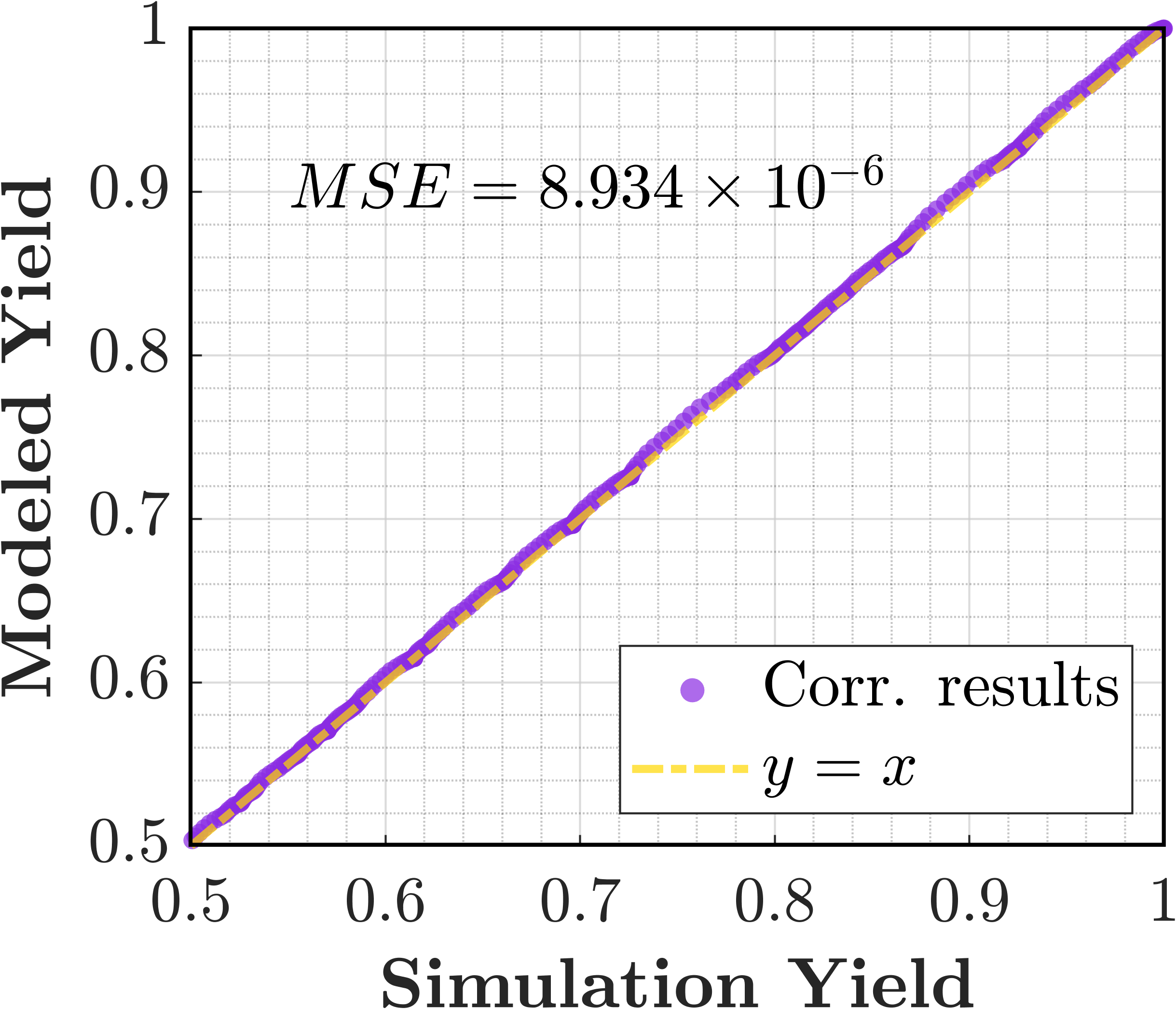}
        \caption{Overlay model.}
        \label{fig:w2w_ovl_correlation}
    \end{subfigure}
    \hfill
    \begin{subfigure}[b]{0.49\linewidth}
        \centering
        \includegraphics[width=\linewidth]{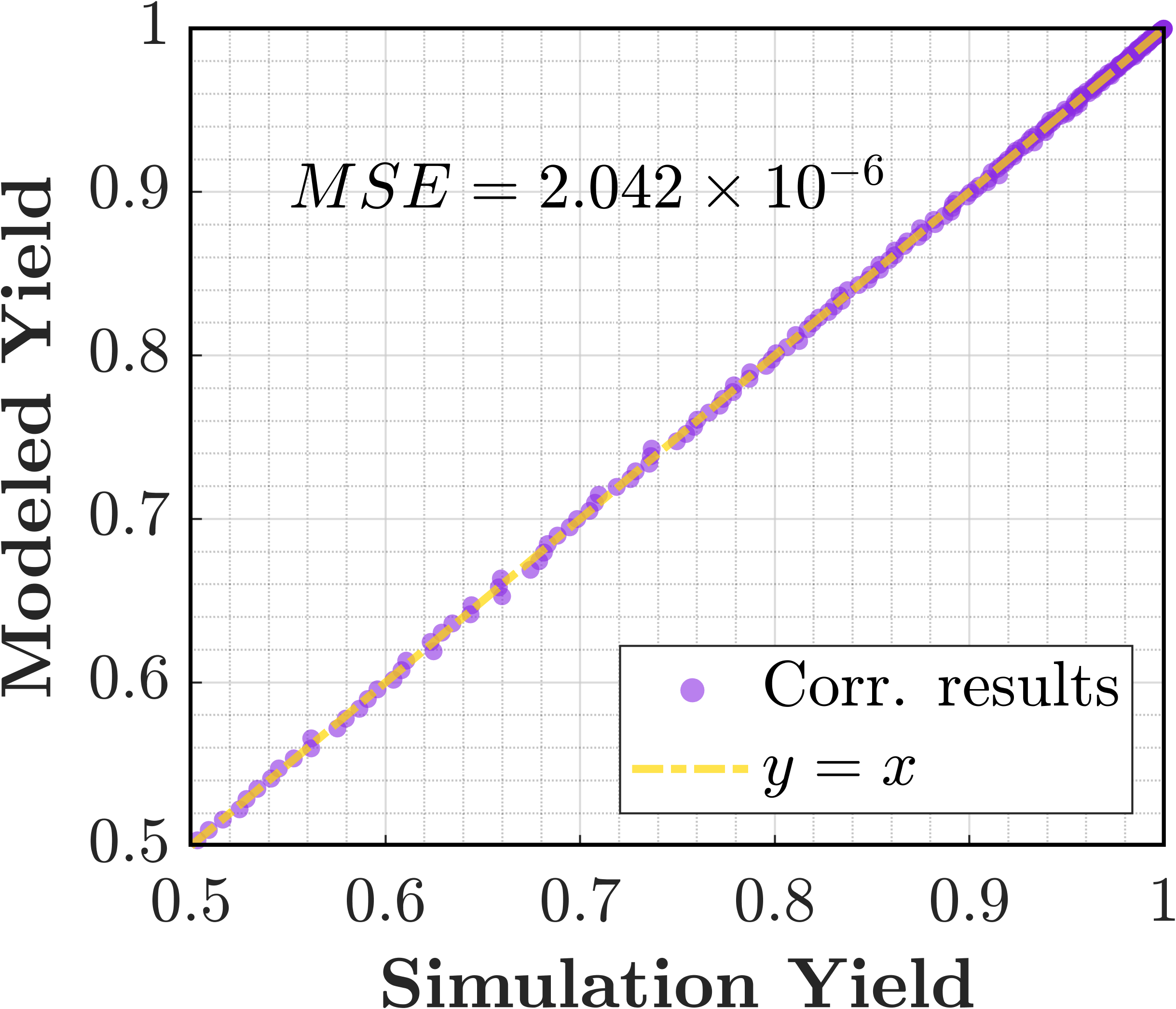}
        \caption{Cu recess model.}
        \label{fig:w2w_ce_correlation}
    \end{subfigure}
    \caption{Correlation results of the overlay model and the Cu recess model with the simulation data for W2W HB.}
    \label{fig:w2w_ovl_ce_correlation}
\end{figure}

We vary input parameters for both the model and simulator, including translation error, rotation error, warpage, die size, pad layouts, and other relevant factors.
The 300 comparison results (purple points) and the mean squared error (MSE) in Fig. \ref{fig:w2w_ovl_correlation} indicate that our model aligns closely with the simulation results, thereby validating its reliability and accuracy.

\subsection{Cu Recess Model}
We can assume the pad height after the CMP process follows a normal distribution according to \cite{Kim2020, De2023}. 
Taking the dielectric surface as the zero reference level, the pad height is considered negative for recessed pads and positive for protruded pads.
It is evident that the combined height of the top and bottom pads also follows a normal distribution. 
Let $h$ denote the sum of heights of two corresponding pads. 
The mean of this distribution is represented by $\mu_h$, and the variance by $\sigma_h^2$. 
To prevent Cu bonding failure and dielectric delamination, the combined pad height $h$ must be constrained within a safe range $(\zeta_-,\zeta_+)$.
Below we discuss the calculation of $\zeta_-$ and $\zeta_+$.

\paragraph{Calculation of $\zeta_-$}
As observed in \cite{De2023, Lin2024, Lin2024observation}, the height variation resulting from Cu expansion during annealing exhibits a linear correlation with the annealing temperature. 
The \textit{lower bound} $\zeta_-$ of the total Cu heights required to form a qualified Cu bonding area is determined by the cumulative Cu expansion after PBA. 
This ensures that any gap between the pads caused by recesses is sufficiently filled with Cu, thereby preventing bonding failure.

\paragraph{Calculation of $\zeta_+$}
The \textit{upper bound} $\zeta_+$ represents the critical condition for dielectric delamination, occurring when the combined pad height reaches this threshold. 
It is important to note that the surface roughness reduces the effective contact area of between the two bonding surfaces, which can exacerbate the risk of delamination. 
To calculate the normalized effective contact area $A_b^*(\sigma_z, R_z, E_d, w)$, we adopt the asperity-based roughness model proposed by \cite{Gui1999,Maugis1996}. 
This model incorporates key roughness and bonding parameters, including the standard deviation of asperity height $\sigma_z$, asperity cap radius $R_z$, Young's modulus of contact surface material $E_d$, and the bonding energy under full contact $w$
\footnote{Modeling the interaction between two rough surfaces requires normalization of both the surface roughness $\sigma_z$ and Young's modulus $E_d$ \cite{Rieutord2006}.}. 
The maximum tolerable peeling stress $\sigma_{tol}$, beyond which delamination may occur, can be expressed as
\begin{equation}
    \sigma_{tol}=A_b^*(\sigma_z, R_z, E_d, w)\times\sqrt{\frac{2E_dw}{t_{d}}}
    \label{eq:max_peeling_stress}
\end{equation}
where $t_d$ represents the thickness of the surface material \cite{Hutchinson1991}. 
As the ambient temperature fluctuates during PBA, the thermal expansion mismatch between the metal and dielectric materials induces various stresses at the bonding interface \cite{Zhao2024}. 
Among various interfaces in the bonding structure, the dielectric-dielectric (e.g. $\mathrm{SiO_2}$-$\mathrm{SiO_2}$) is more susceptible to delamination due to its relatively lower bonding strength and the elevated peeling stress observed at the end of the annealing dwell stage \cite{Ji2020,Fujii2023}. 
For simplification, we employ a fitting model to evaluate the dependence of dielectric interface peeling stress, based on the asperity and bonding parameters.
\begin{equation}
    \sigma_{peel}=k_{peel}\cdot D_{Cu}\cdot (h-h_0)
    \label{eq:linear_peeling_stress_model}
\end{equation}
where $D_{Cu}$ represents the Cu pattern density, $h_0,k_{peel}$ are fitting parameters, and $k_{peel}$ is influenced by factors such as annealing temperature, pad shape, pad arrangement, pad structure, etc. \cite{Ji2020,Zhao2024,Wang2023}.
To avoid delamination, one should have 
\begin{equation}
\sigma_{tol}\ge\sigma_{peel}\Rightarrow h\le h_{peel}    
\end{equation}
Additionally, since the Cu protrusion after CMP can lead to delamination, the \textit{upper bound} of the combined pad height is expressed by
\begin{equation}
    \zeta_+=\min\{0,h_{peel}\}
    \label{eq:upper_bound_height_sum}
\end{equation}
To summarize, the POS of this pad during PBA is given by 
\begin{equation}
    POS_{cr,pad}=\frac{1}{\sqrt{2 \pi\sigma_h^2}} \int_{\zeta_-}^{\zeta_+} e^{-\frac{(h-\mu_h)^2}{2 \sigma_h^2}}\text{d}h
    \label{eq:bulge_out_pos_pad}
\end{equation}
Assume that a die contains $N_{cr}$ critical pads and $N_r$ groups of non-power/ground redundant pads, with each group containing $M_r$ replicas.
The POS of the critical pads is given by $POS_{cr,pad}^{N_{cr}}$.
The POS of the redundant pads is expressed as $[1-(1-POS_{cr,pad})^{M_r}]^{N_r}$.
The die yield, regarding Cu recess variations, is given by the product of these two terms.
\begin{equation}
\begin{aligned}
   Y_{cr,W2W}&=POS_{cr,die}\\
   &=POS_{cr,pad}^{N_{cr}}\cdot \left[1-\left(1-POS_{cr,pad}\right)^{M_r}\right]^{N_r}
    \label{eq:yield_ce_W2W}
\end{aligned}
\end{equation}

We vary input parameters of Cu recess, pitch, roughness, etc. to validate the Cu recess model.
Fig. \ref{fig:w2w_ce_correlation} presents the correlation between the model predictions and simulation results.

\subsection{Defect Model} 
\begin{figure}
    \centering
    \includegraphics[width=1\linewidth]{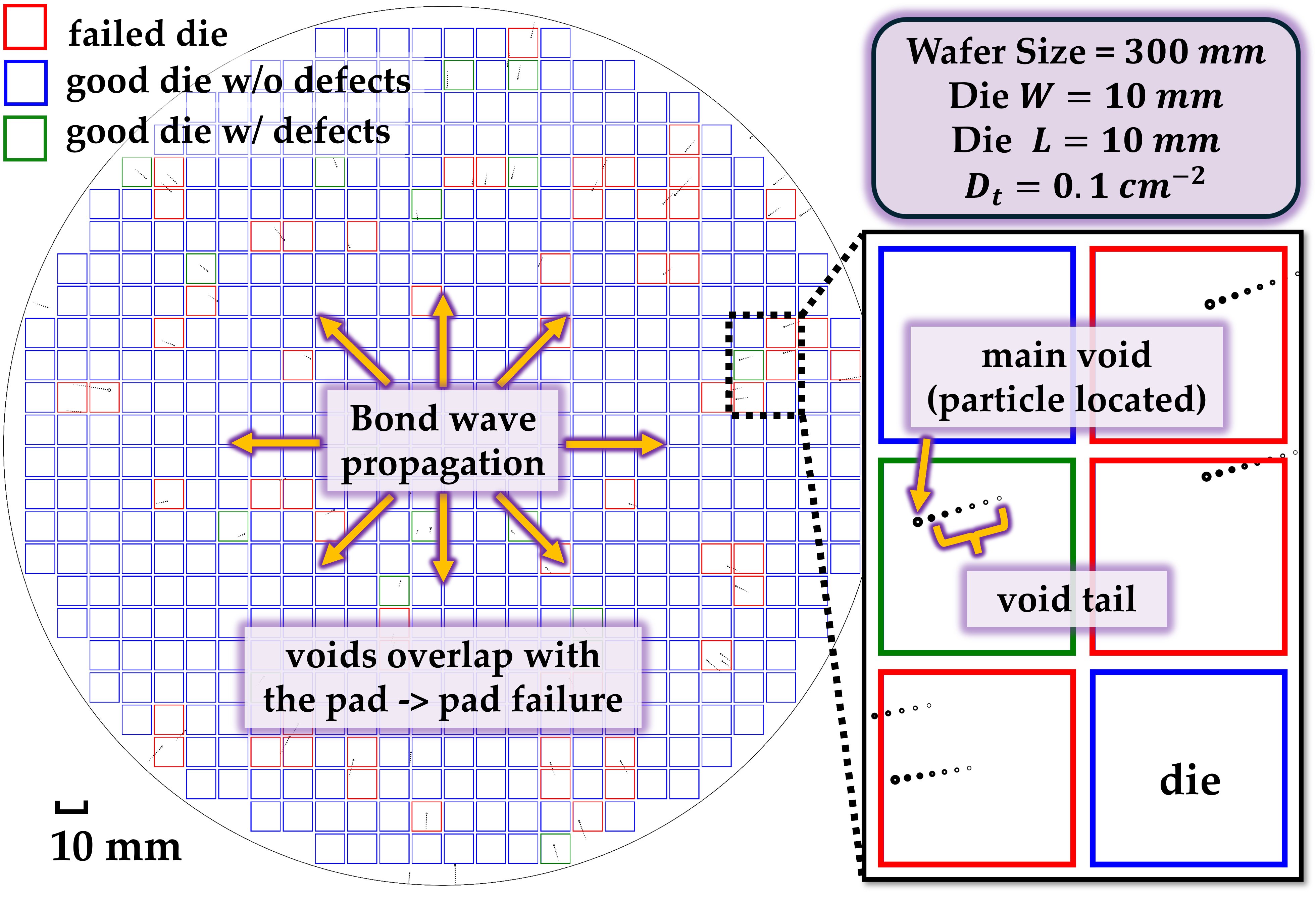}
    \caption{Visualization of the void formation simulation.}
    \label{fig:void_visualizer}
\end{figure}
The relationship between a particle's properties and resulting void size is complex, involving factors such as particle thickness, Young's modulus, and wafer adhesion energy. 
Due to the significant discrepancies between theoretical predictions and experimental observations in \cite{Nagano2022}, it is more practical to develop a fitting model that estimates the void size based on certain process information. 
Furthermore, as demonstrated in \cite{Lau2023}, for particles of specific size and material, both the main void size and the void tail length exhibit linear correlations with the particle’s location and the square root of its thickness.
To capture these relationships, we adopt simple linear models, with fitting parameters derived from the slope and intercept of the observed trends reported in \cite{Nagano2022}.

\subsubsection{Defect Shape Modeling}
We model the size $r_{mv}$ of the main void located at a distance $L$ from the wafer center, where $0\le L<R$, using the following relationship:
\begin{equation}
    r_{mv} = (k_rL+k_{r_0}){t}^{1/2}
    \label{eq:void_size_model}
\end{equation}
where $t$ denotes the particle thickness. Similarly, the void tail length $l$ can be modeled by
\begin{equation}
    l=k_lL{t}^{1/2}
    \label{eq:void_tail_length_model}
\end{equation}
where $k_r,k_{r_0},k_l$ are fitting parameters.
Fig. \ref{fig:void_visualizer} visualizes the simulated void formation, which closely resembles the scanning acoustic microscopy images of voids reported in \cite{Nagano2022}.
Since the average void tail length on the wafer can reach a few millimeters, more than 10 times the scale of the main void size (typically a few hundred \SI{}{\micro\meter}) in W2W HB, the defect geometry can be reasonably simplified as a straight line characterized by its length $l$ and outward orientation $\theta$. 
Furthermore, a die is considered to have failed if the void tail overlaps the functional pad array area, since the void size is typically much larger than the HB pitch ($\le$ 10 $\mathrm{\mu m}$).

We assume the thickness distribution of particle defects as $D(t)$. A typical form of $D(t)$ can be modeled as \cite{Glang1991}
\begin{equation}
D(t)=D_t\cdot\frac{(z-1)\cdot t_0^{z-1}}{t^{z}}, \qquad t>t_0
\label{eq:defect_thickness_distribution}
\end{equation}
where $t_0$ denotes the minimum particle thickness, and $D_t$ represents the total particles count per unit area across all thicknesses. 
The parameter $z$ controls the shape of the distribution and empirically ranges between 2 and 3 \cite{Stapper1984, Bruls1992}.  
The parameters in this distribution are obtained by fitting the model to the data of cleanroom concentration of particles presented in \cite{Dylan2024}.
By Eq. \ref{eq:void_tail_length_model}, \ref{eq:defect_thickness_distribution}, the distribution of void tail length can be calculated by
\begin{equation}
\label{eq:void_tail_length_distribution}
    f_l(l)=
    \left\{
    \begin{aligned}
        &\frac{2D_t(z-1)l}{zk_l^2R^2t_0}, \qquad l\le k_lRt_0^{1/2}\\
        &\frac{2D_t(z-1)(k_l^2R^2t_0)^{z-1}}{zl^{2z-1}}, \qquad l>k_lRt_0^{1/2}
    \end{aligned}
    \right.
\end{equation}
where $R$ represents the wafer radius. 
The comparison of the derived $f_l(l)$ and the simulated distribution is shown in Fig. \ref{fig:void_tail_length_distribution}, confirming the precision of the derivation.

\subsubsection{Dilation-based Critical Area Calculation}
\begin{figure}[t]
    \centering
    \includegraphics[width=1\linewidth]{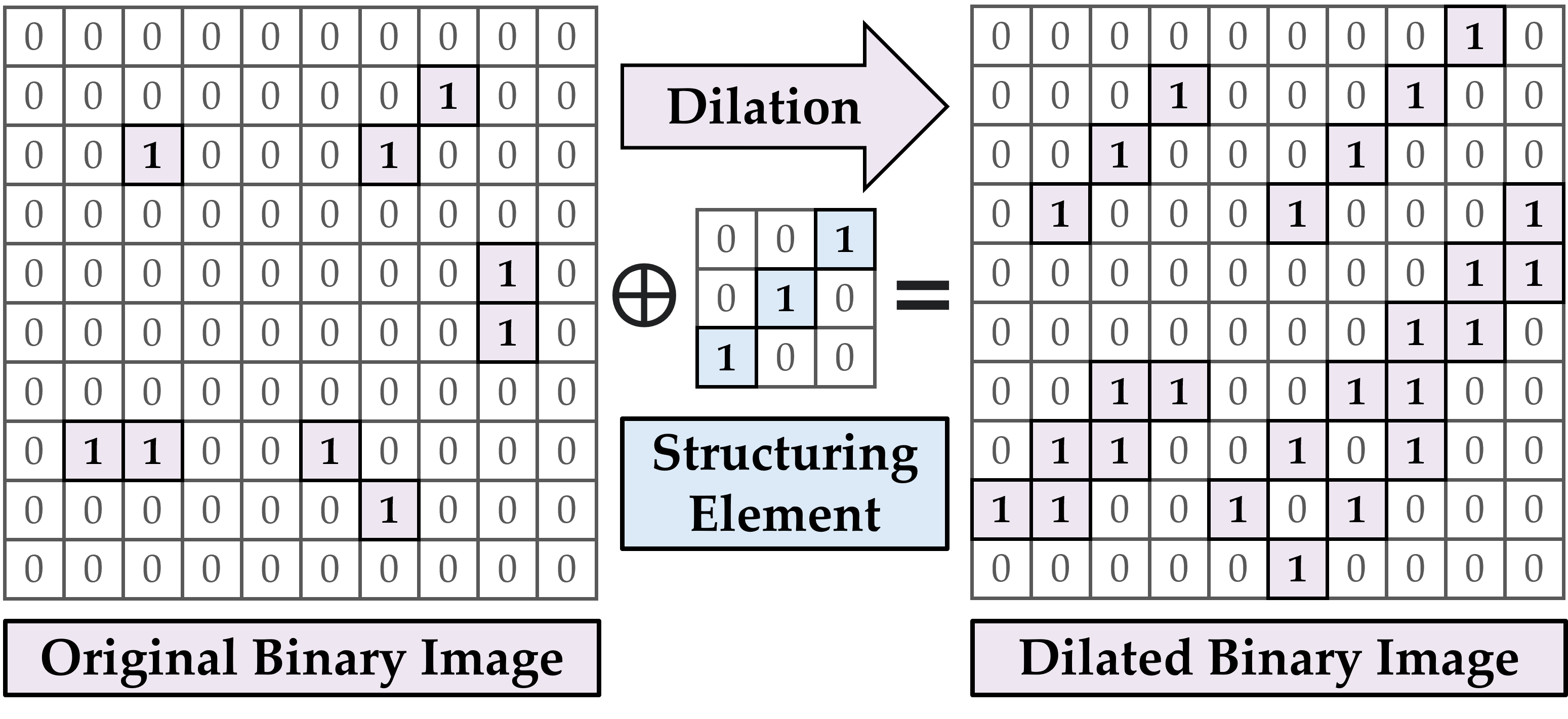}
    \caption{An example of a dilation operation on a binary image using a structuring element. The structuring element defines the shape and extent of expansion applied to the original binary image. In this context, dilation simulates how a defect (void) affects nearby pads during hybrid bonding.}
    \label{fig:dilation_example}
\end{figure}
\begin{figure*}[t]
    \centering    \includegraphics[width=\textwidth]{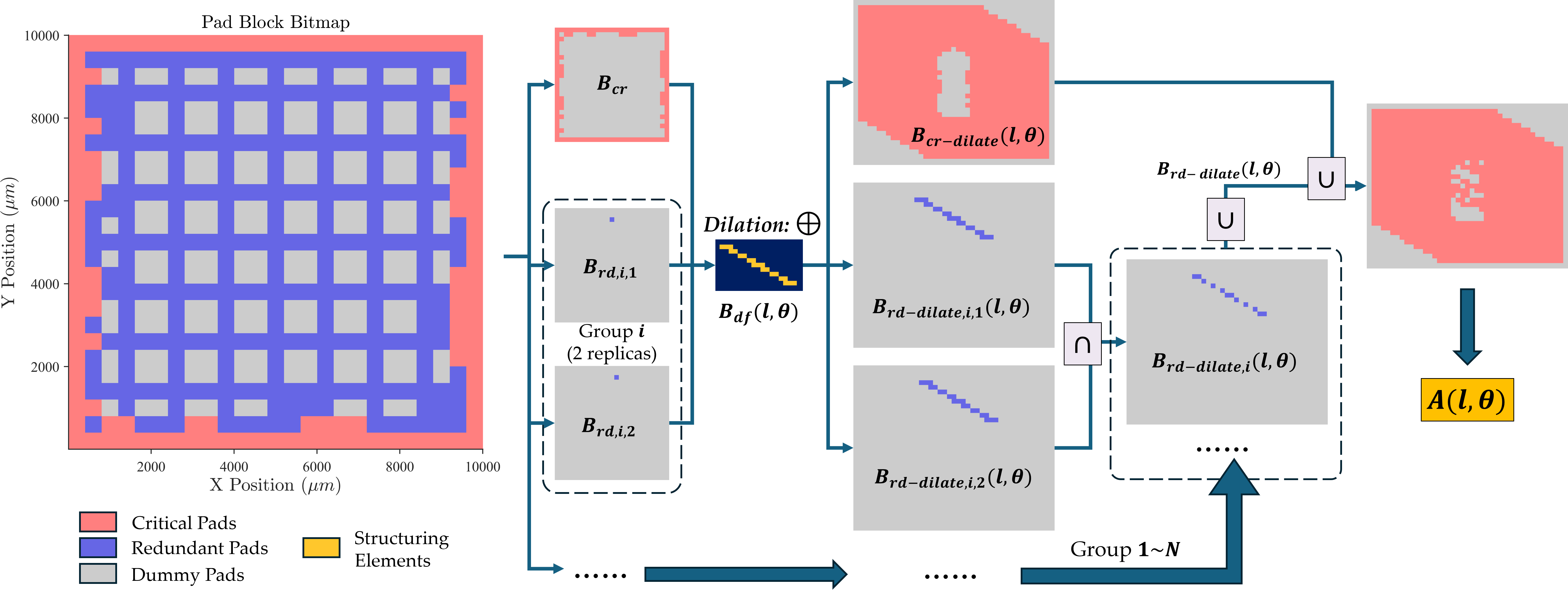}
    \caption{Dilation-based critical area calculation diagram. In this example, the die size is \SI{10}{\milli\meter}$\times$\SI{10}{\milli\meter}, and the gridding resolution in modeling is \SI{400}{\micro\meter}$\times$\SI{400}{\micro\meter}. The die consists of 20\% critical pads, 50\% redundant pads, and 30\% dummy pads. The structuring element is generated based on the length $l$ and the orientation $\theta$ of a void tail and the gridding resolution.}
    \label{fig:dilation_method}
    \vspace{-10pt}
\end{figure*}
Die failure can result from the failure of a critical pad or from the simultaneous failure of a redundant pad and all its replicas. 
Based on this relationship, the critical area of the die is defined as the region where a defect would trigger either of these failure mechanisms.
When modeling designs with specific pad layouts, including the location and number of different types of I/O pads, such as critical, redundant, and dummy pads, deriving such analytical expressions of critical area for random layouts is not straightforward.
To cope with the challenge, we propose a \textit{dilation-based} method for the critical area calculation.
\textit{Dilation}, a fundamental morphological operation, expands the boundaries of geometric features through a predefined structuring element. 
Intuitively, it can be understood as a process that `grows' or `thickens' objects in a binary image \cite{Gonzalez2009}. 
An example of a dilation operation on a binary image is shown in Fig. \ref{fig:dilation_example}.
This concept can be effectively applied to model the potential impact area of defects on a die layout, regardless of the shape of either the defect or the layout.

The process of dilation-based critical area calculation is illustrated in Fig. \ref{fig:dilation_method}.
To enable efficient computation, the die is divided into a grid of pad blocks, with each pad block containing only one type of I/O pad.
Each pad block is treated as an atomic unit in the yield modeling, meaning that all pads within a block either survive or fail together.
Using a finer gridding resolution enables a more accurate representation of the layout, but at the cost of increased computational time both in modeling and simulation, as it leads to larger bitmap dimensions and consequently longer dilation operations.
To ensure acceptable accuracy, the gridding resolution must be carefully selected.
This choice is typically guided by the size of the defects under consideration.
The relationship between gridding resolution and defect dimensions will be discussed in detail in Section IV.
Note that redundant pads and their corresponding replicas are not necessarily located within the same pad block.
Based on this gridding, we generate a pad block bitmap for each type of I/O pad.
Similarly, a void tail defect is also represented as a bitmap by using Bresenham's line algorithm \cite{Bresenham1965}, constructed according to its length and orientation.
For critical pad blocks, the critical area bitmap $\mathbf{B}_{cr-dilate}$ is obtained by dilating the critical pad block bitmap $\mathbf{B}_{cr}$ using the defect bitmap $\mathbf{B}_{df}(l,\theta)$, as defined by
\begin{equation}
    \mathbf{B}_{cr-dilate}(l,\theta)=\mathbf{B}_{cr}\oplus \mathbf{B}_{df}(l,\theta)
    \label{eq:critical_area_bitmap_for_critical_pad_blocks}
\end{equation}
where $\oplus$ is the dilation operator, and $\mathbf{B}_{df}(l,\theta)$ is the approximated bit map of a void tail defect of length $l$ and orientation $\theta$. 
For redundant pad blocks, suppose a die contains $N_{blk}$ groups of redundant pad blocks, within each group consists of one main pad block and its corresponding replicas.
The HB process typically requires Class 1 / ISO 3 cleanrooms and equipment or better, which results in a relatively low particle defect density (\SI{\sim 0.1}{\per\centi\meter\squared} \cite{Dylan2024}).
The probability of multiple particle defects occurring in close proximity and independently disabling both redundant replicas is extremely low, particularly given that these replicas are typically placed near one another.
As a result, the critical area for each group of redundant pads can be reasonably approximated as the region within which a single defect is sufficient to cause the failure of all replicas.
Assuming that the $i$-th group contains $M_r$ replicas, the critical area bitmap of this group $\mathbf{B}_{rd-dilate,i}$ can be approximated by the intersection of the dilated bitmaps of all redundant replica pad blocks $\mathbf{B}_{rd,i,1}\sim \mathbf{B}_{rd,i,M_r}$, where each is individually dilated using the defect bitmap.
Formally, this is expressed as
\begin{equation}
\begin{aligned}
    \mathbf{B}_{rd-dilate,i}(l,\theta)&=\bigcap_{j=1}^{M_r}(\mathbf{B}_{rd,i,j}\oplus \mathbf{B}_{df}(l,\theta))\\
\end{aligned}
\label{eq:critical_area_bitmap_for_redundant_pad_blocks}
\end{equation}
Finally, the critical area of the die, $A(l,\theta)$, regarding a void tail defect of length $l$ and orientation $\theta$, is defined as the area of the union of the two bitmaps $\mathbf{B}_{cr-dilate}$ and $\mathbf{B}_{rd-dilate}$.
Here, $\mathbf{B}_{rd-dilate}$ represents the union of all group-wise redundant pad block bitmaps after dilation, i.e., $\mathbf{B}_{rd-dilate,1}\sim \mathbf{B}_{rd-dilate,N_{blk}}$.
This can be expressed as
\begin{equation}
\begin{aligned}
    A(l,\theta)&\approx Area(\mathbf{B}_{cr-dilate}(l,\theta)\cup \mathbf{B}_{rd-dilate}(l,\theta))\\
    &=Area(\mathbf{B}_{cr-dilate}(l,\theta)\cup(\bigcup_{i=1}^{N_{blk}} \mathbf{B}_{rd-dilate,i}(l,\theta)))
\end{aligned}
\label{eq:critical_area_of_critical_and_redundant_pad_blocks}
\end{equation}
where $Area(\cdot)$ denotes the area of the region represented by the bitmap.
Hence, the average number of particle-induced void tail defects that will cause a die to fail, $\Lambda$, can be expressed as
\begin{equation}
    \begin{aligned}
        \Lambda=\int_0^\infty\int_0^{2\pi} A(l,\theta)f_l(l)\mathrm{d}\theta\mathrm{d}l
    \end{aligned}
    \label{eq:avg_fatal_defects}
\end{equation}
Using the Poisson yield model \cite{Koren1998}, the yield with respect to the particle-induced void formation is given by
\begin{equation}
\begin{aligned}
    Y_{df,W2W}=\exp(-\Lambda)
\end{aligned}
\label{eq:yield_defect_W2W}
\end{equation}
Compared to previous analytical computation, the dilation-based modeling method is more computationally expensive due to the repeated processing of binary images, although the use of the gridding strategy helps mitigate this overhead.
However, for a given pad layout, the critical area only needs to be computed once. 
The result can be stored in a look-up table and retrieved as needed when varying the process parameters, thereby mitigating the drawback of increased computation time.
To summarize, compared to traditional analytical models, the dilation-based approach provides improved geometric accuracy for arbitrary layouts and defect shapes, along with better scalability for large and complex designs, while its computational overhead has minimal impact during the application phase.

\begin{figure}[b]
    \centering
    \begin{subfigure}[b]{0.48\linewidth}
    \centering
    \includegraphics[width=\linewidth]{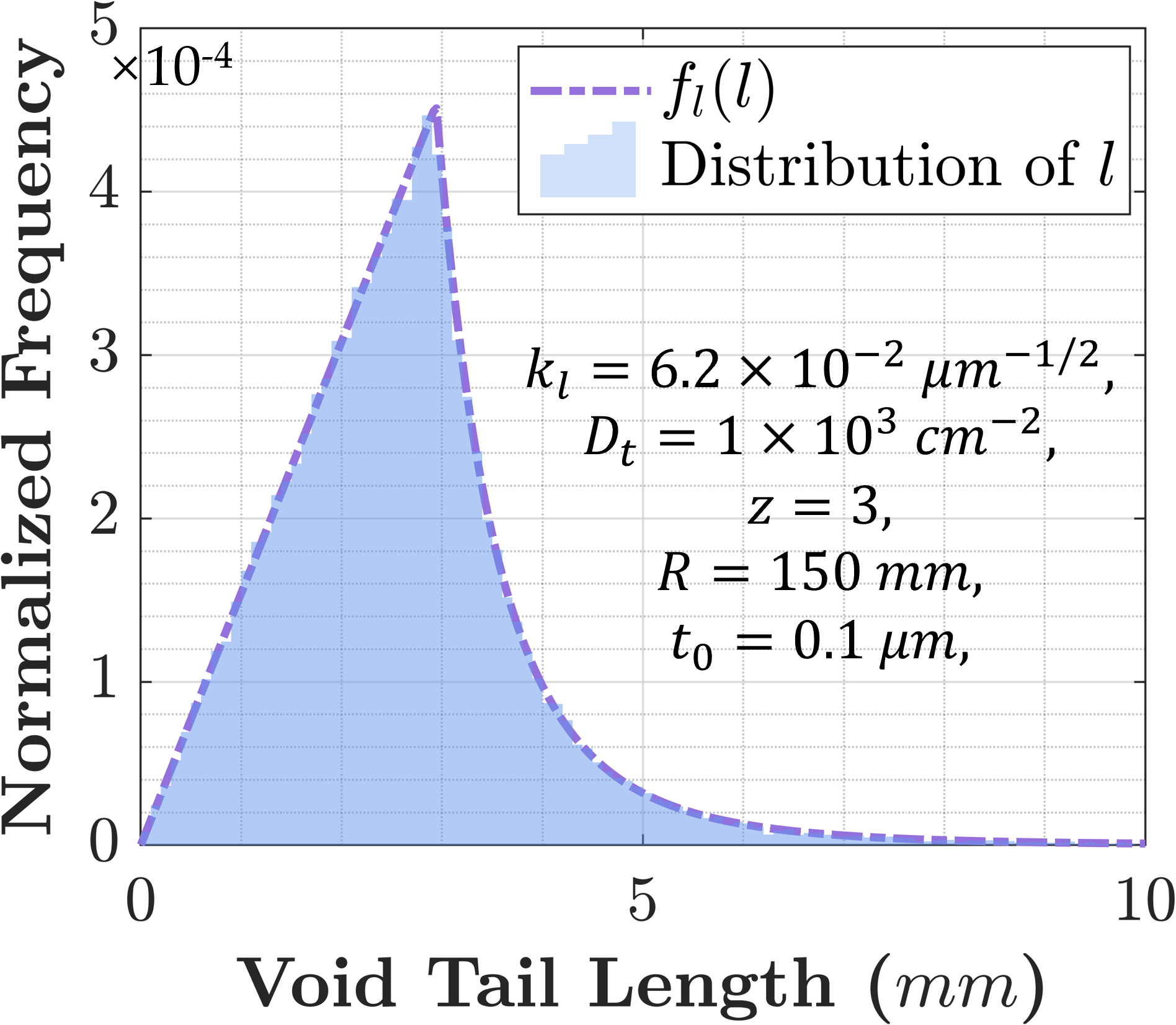}
    \caption{Void tail length distribution.}
    \label{fig:void_tail_length_distribution}
    \end{subfigure}
    \hfill
    \begin{subfigure}[b]{0.49\linewidth}
        \centering
        \includegraphics[width=\linewidth]{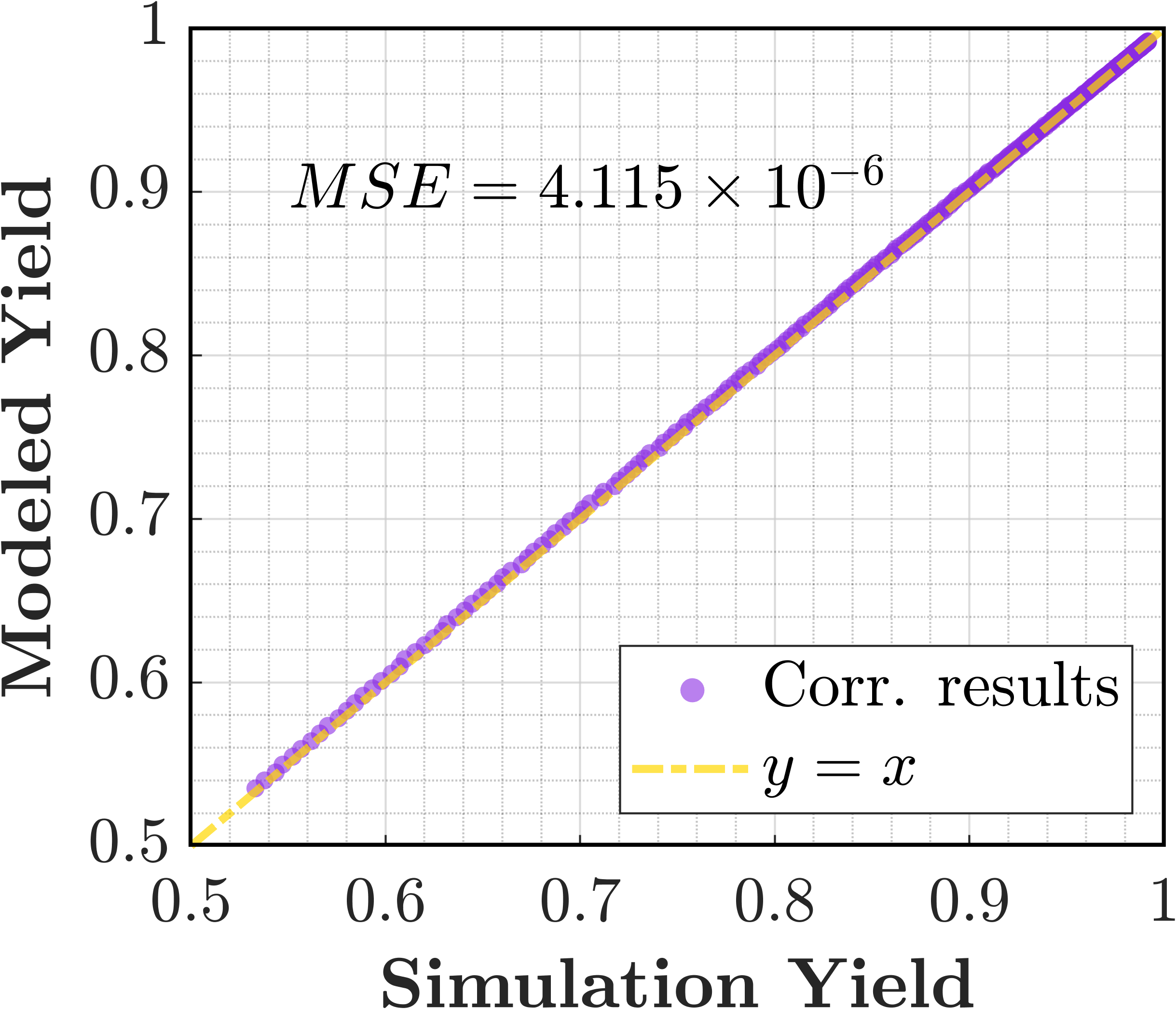}
        \caption{Defect model.}
        \label{fig:w2w_pd_correlation}
    \end{subfigure}
    \caption{Correlation results of the void tail length distribution and the defect model with the simulation data for W2W HB.}
    \label{fig:defect_model_correlation}
\end{figure}

We vary the input parameters of particle defect density, die size, wafer size, etc., to validate the defect model. Fig. \ref{fig:w2w_pd_correlation} demonstrates the correlation of the defect yield with the simulation results.

\subsection{Overall Bonding Yield Model}
To develop the overall bonding yield model for W2W hybrid bonding, we assume that the overlay error, Cu recess variations, and particle defects affect die yield independently. By integrating the individual yield components, namely Eq. \ref{eq:yield_ovl_W2W}, \ref{eq:yield_ce_W2W}, \ref{eq:yield_defect_W2W}, the assembly yield is
\begin{equation}
\begin{aligned}
Y_{W2W}& = Y_{ovl,W2W}\cdot Y_{cr,W2W}\cdot Y_{df,W2W}
\end{aligned}
\label{eq:yield_asbly_W2W}
\end{equation}

\subsection{D2W Hybrid Bonding Yield Models}
We extend the yield model to the D2W HB scenario. It is assumed that the Cu expansion behavior during PBA, as observed in W2W bonding, remains applicable to D2W HB. However, the yield components associated with overlay error and particle defects must be revised to reflect the distinct characteristics of the D2W bonding process.
\subsubsection{Overlay Model}
In D2W hybrid bonding, systematic overlay errors occur independently for each die.
Due to the smaller die size, an identical marker misalignment at the die edge leads to larger rotation $\alpha$ and magnification $E$ errors compared to the W2W case.
Similar to W2W HB, the overall POS for a die in D2W bonding is determined by the minimum POS among all interconnection pads, excluding dummy pads and power/ground I/O pads.
For a die with $N_{cr}$ critical and $N_{rd}$ redundant pads, its overlay yield can be written as
\begin{equation}
\begin{aligned}
    Y_{ovl,D2W}=\frac{1}{\sigma_1 \sqrt{2 \pi }} \min_{i\in[1,N_{cr}+N_{rd}] }\left\{\int_{-\delta-s_i}^{\delta-s_i} e^{-\frac{u^2}{2 \sigma_1^2}}\text{d}u\right\}   
\end{aligned}
\label{eq:yield_ovl_D2W}
\end{equation}

\subsubsection{Defect Model}
\begin{figure}[t]
    \centering
    \begin{subfigure}[b]{0.50\linewidth}
        \centering
        \includegraphics[width=\linewidth]{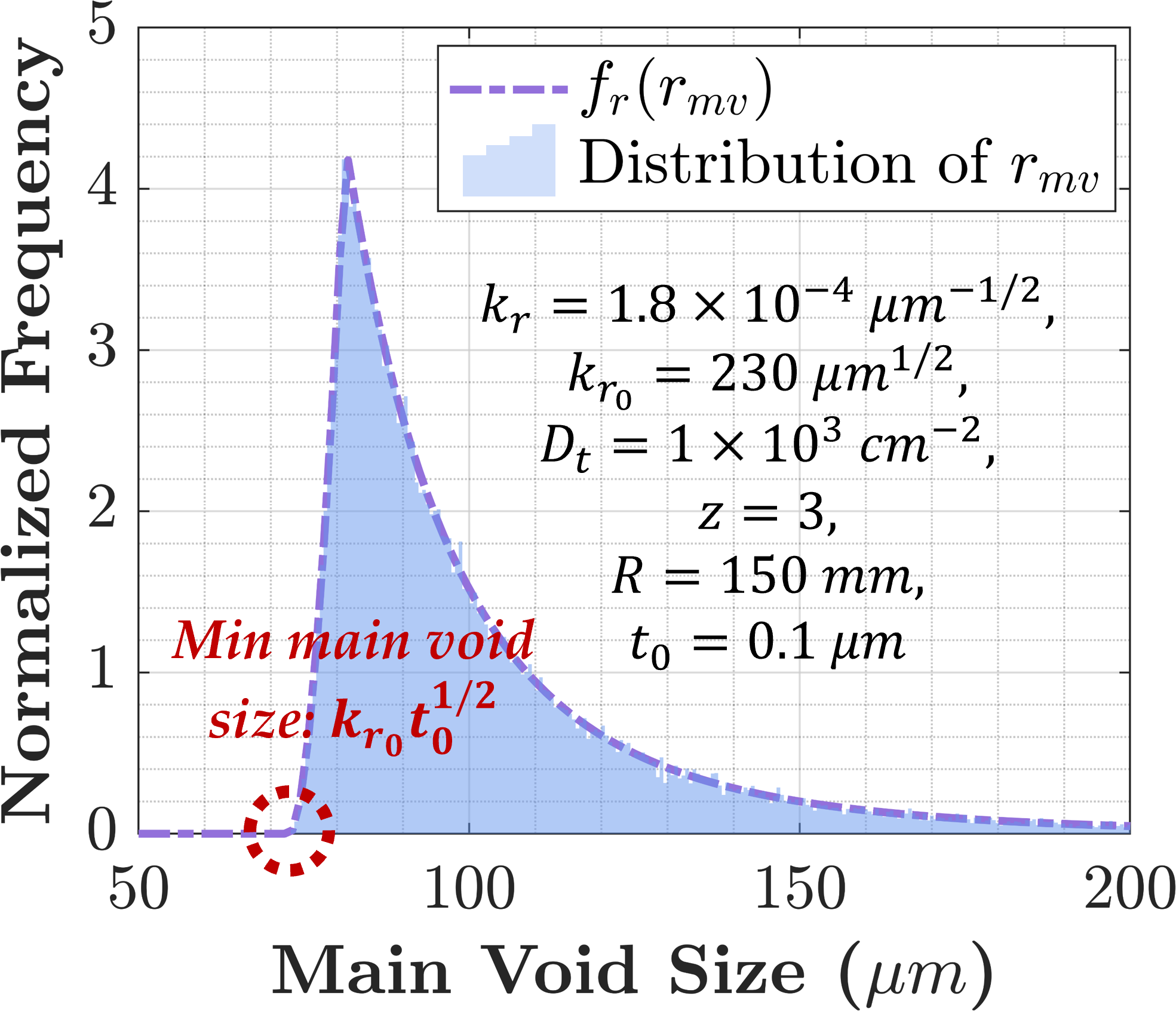}
        \caption{Main void size distribution.}
        \label{fig:main_voids_distribution}
    \end{subfigure}
    \hfill
    \begin{subfigure}[b]{0.48\linewidth}
        \centering
        \includegraphics[width=\linewidth]{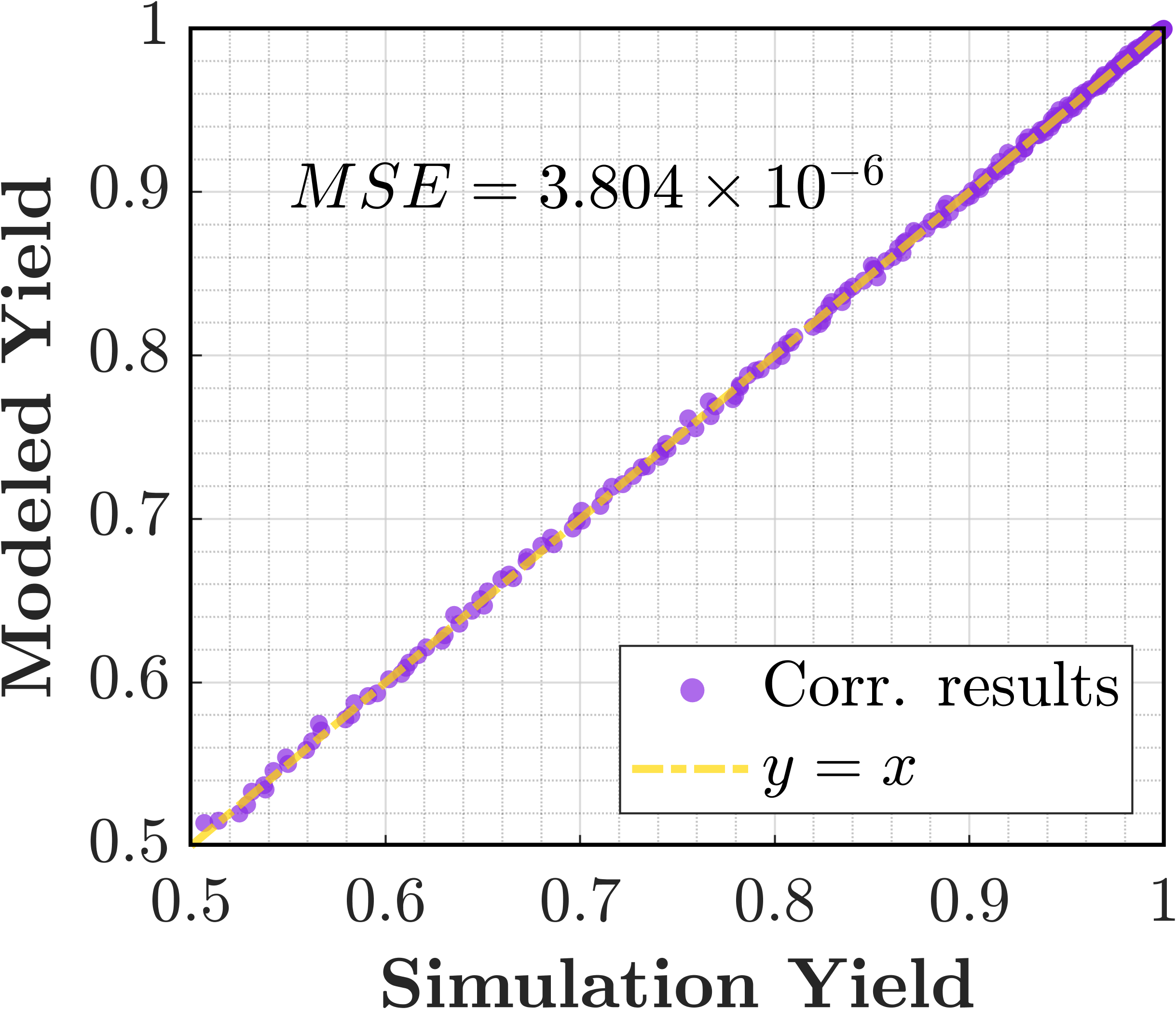}
        \caption{Overlay model.}
        \label{fig:d2w_ovl_correlation}
    \end{subfigure}

    \begin{subfigure}[b]{0.49\linewidth}
        \centering
        \includegraphics[width=\linewidth]{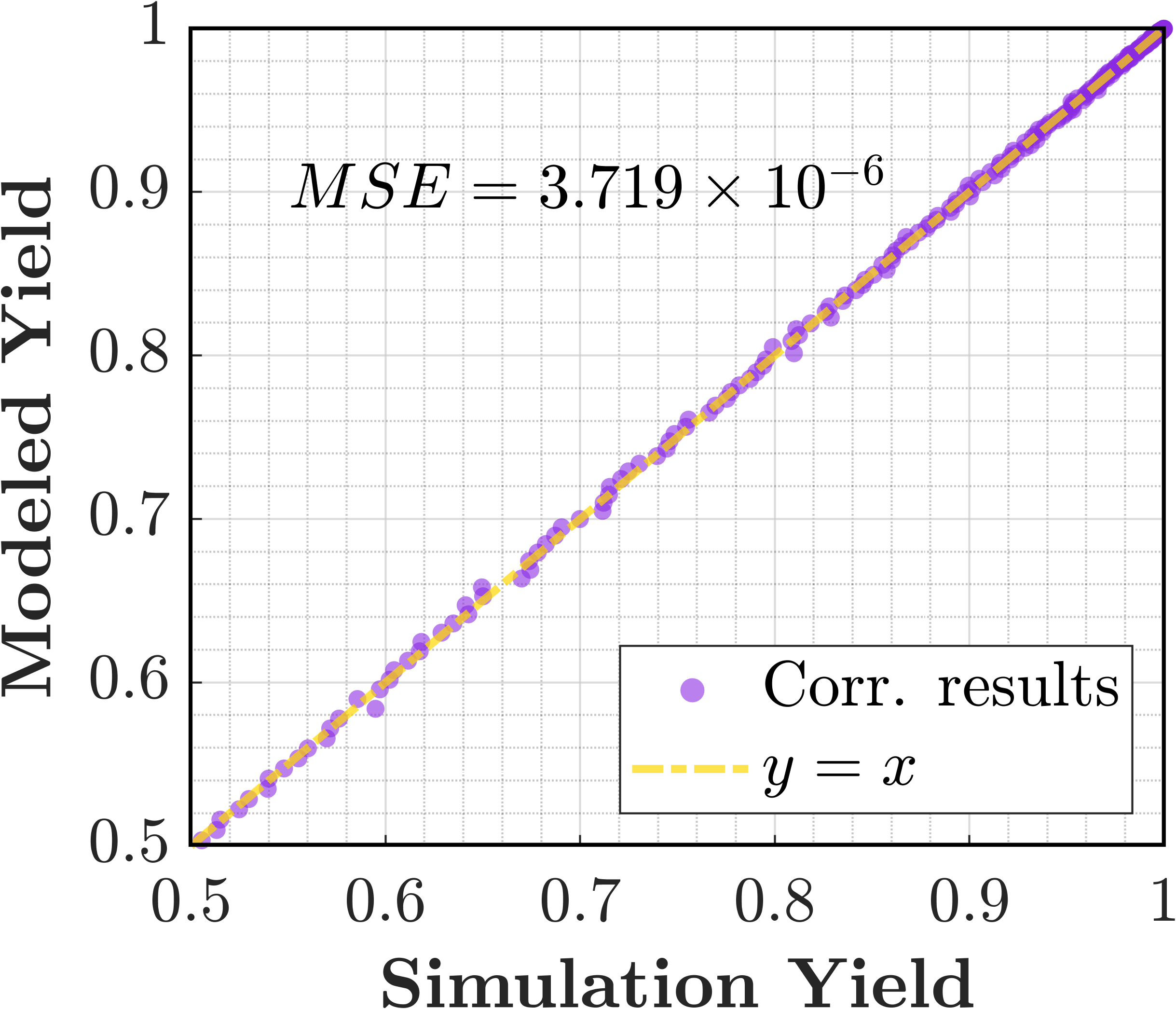}
        \caption{Cu recess model.}
        \label{fig:d2w_ce_correlation}
    \end{subfigure}
    \hfill
    \begin{subfigure}[b]{0.49\linewidth}
        \centering
        \includegraphics[width=\linewidth]{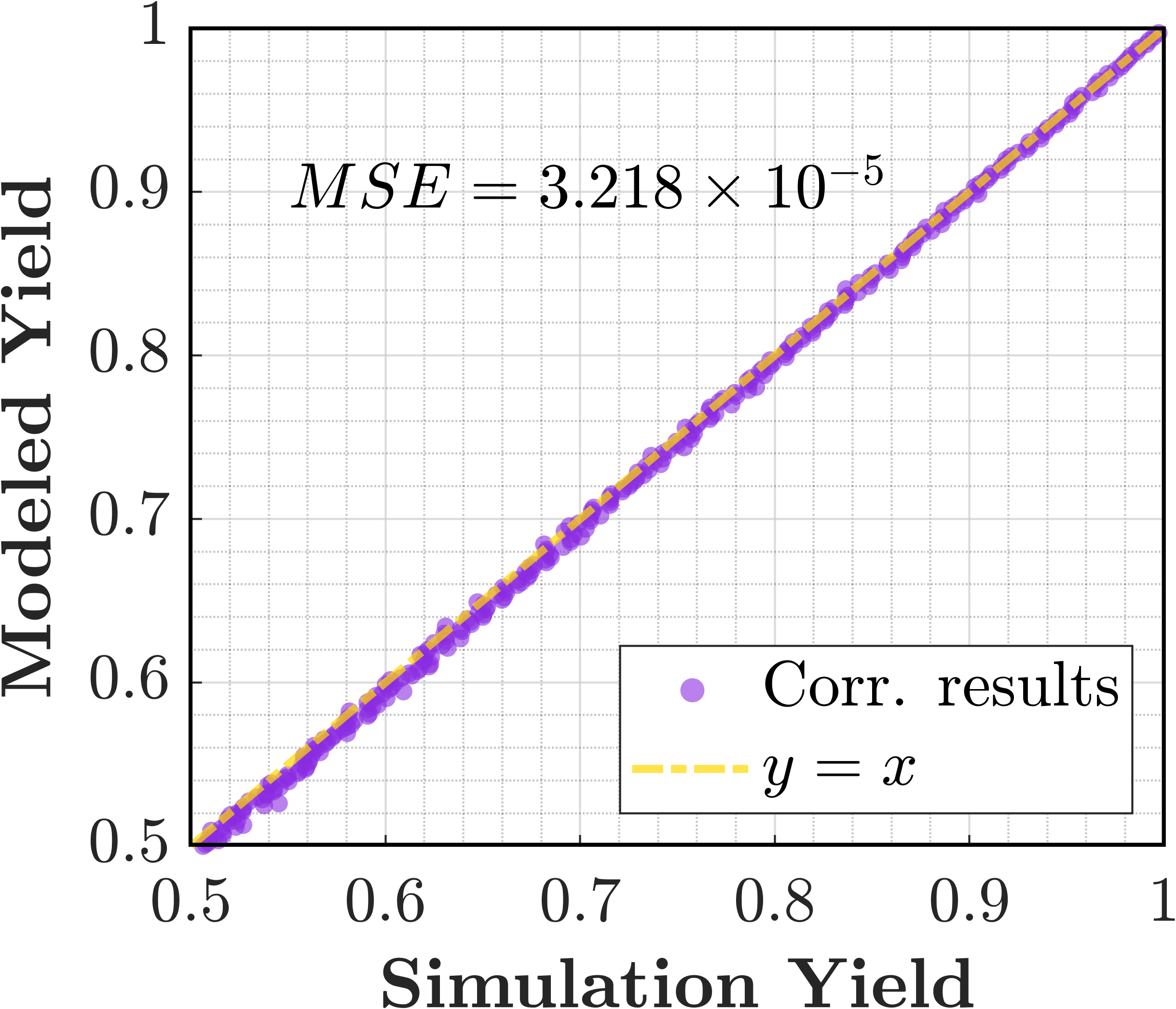}
        \caption{Defect model.}
        \label{fig:d2w_pd_correlation}
    \end{subfigure}
    \caption{Correlation results of main void size distribution and the yield model with the simulation data for D2W HB.}
    \label{fig:d2w_yield_term_correlation}
\end{figure}

Given the smaller die scale in D2W HB, void tail formation is unlikely to occur.
Thus, the D2W defect model considers only main void-induced failures.
By combining Eq. \ref{eq:void_size_model}, \ref{eq:defect_thickness_distribution}, the probability density function (PDF) of the main void size $r_{mv}$ can be given by 
\begin{equation}
\small
    \begin{aligned}
    f_r(r_{mv})=
    \left\{
    \begin{aligned}
    &\frac{D_t(z-1)t_0^{z-1}}{k_r^2R^2}\times[\frac{2r_{mv}}{zt_0^z}+\frac{2k_{r_0}^{2z}}{z(2z-1)r_{mv}^{2z-1}}\\
    &-\frac{2k_{r_0}}{(z-\frac{1}{2})t_0^{z-\frac{1}{2}}}], \quad k_{r_0}t_0^{1/2}<r<(k_rR+k_{r_0})t_0^{1/2},\\
    &\frac{2D_t(z-1)t_0^{z-1}(k_rR+k_{r_0})^{2z-2}}{r_{mv}^{2z-1}}-\\
    &\frac{2D_t(z-1)^2t_0^{z-1}}{k_r^2R^2r_{mv}^{2z-1}}\times
    [\frac{(k_rR+k_{r_0})^{2z}-k_{r_0}^{2z}}{z}\\
    &-\frac{2k_{r_0}(k_rR+k_{r_0})^{2z-1}-2k_{r_0}^{2z}}{z-\frac{1}{2}}+\\
    &\frac{k_{r_0}^2(k_rR+k_{r_0})^{2z-2}-k_{r_0}^{2z}}{z-1}],\quad r\ge(k_rR+k_{r_0})t_0^{1/2}.
    \end{aligned}
    \right.
    \end{aligned}
    \label{eq:main_void_size_distribution}
\normalsize
\end{equation}
where $R$ is the effective radius of the die, i.e., $R=(ab/\pi)^{1/2}$, aiming to remain the average number of particles on the die.
The close alignment between $f_r(r_{mv})$ and the simulated distribution is shown in Fig. \ref{fig:main_voids_distribution}.
Similarly, the dilation-based method is applied for critical area calculation.
The main void is pixelized, and the structuring element is constructed based on its size.
The critical area of the die, $A(r_{mv})$, regarding a main void defect of radius $r_{mv}$, is defined as the area of the union of the two bitmaps $\mathbf{B}_{cr-dilate}$ and $\mathbf{B}_{rd-dilate}$.
This can be expressed as
\begin{equation}\small
A(r_{mv}) \approx Area(\mathbf{B}_{cr-dilate}(r_{mv})\cup \mathbf{B}_{rd-dilate}(r_{mv}))
   \label{eq:main_void_critical_area}
   \normalsize
\end{equation}
By Eq. \ref{eq:main_void_size_distribution}, \ref{eq:main_void_critical_area}, the average number $\Lambda$ of particle-induced main void defects that will fail a die is given by
\begin{equation}
    \begin{aligned}
        \Lambda=\int_{k_{r_0}t_0^{1/2}}^\infty A(r_{mv})f_r(r_{mv})\mathrm{d}r_{mv}
    \end{aligned}
\end{equation}
Using the Poisson yield model \cite{Koren1998}, the yield with respect to the particle-induced void formation for D2W HB is given by
\begin{equation}
\begin{aligned}
    Y_{df,D2W}=\exp(-\Lambda)
\end{aligned}
\label{eq:yield_defect_D2W}
\end{equation}

\subsubsection{Overall Bonding Yield Model}
Similarly, we assume the overlay error, Cu recess variations, and particle defects have independent impacts on the die yield for D2W HB. Fig. \ref{fig:d2w_ovl_correlation}, \ref{fig:d2w_ce_correlation}, \ref{fig:d2w_pd_correlation} show the correlation results of three yield terms, respectively.
By combining Eq. \ref{eq:yield_ce_W2W}, \ref{eq:yield_ovl_D2W}, \ref{eq:yield_defect_D2W}, the bonding yield is by
\begin{equation}
\begin{aligned}
Y_{D2W} = Y_{ovl,D2W}\cdot Y_{cr,D2W}\cdot Y_{df,D2W}
\end{aligned}
\label{eq:yield_asbly_D2W}
\end{equation}

\section{YAP+ Simulator and Model Validation}
To validate the derived model, a Monte Carlo simulator is developed for Cu-$\mathrm{SiO_2}$ HB process. 
Key inputs include the wafer/die information, overlay distribution, Cu recess variations, particle size distribution, model fitting parameters, and failure criteria. 
The similarity of failure mechanisms reported in the Cu-$\mathrm{SiCN}$ hybrid bonding process further supports the general applicability of the YAP+ modeling framework \cite{Nagano2023, Zhang2024, Kim2020}.
With appropriate configuration of the input parameters, the model can be readily adapted to various bonding scenarios.
Table \ref{tab:model_parameters} presents the baseline model parameters, with additional details available in our code. 
These are the parameter values used in our experiments unless otherwise stated.  
Fig. \ref{fig:simulation_workflow} outlines the simulation workflow. 

\begin{table}[t]
\centering
\small
\begin{threeparttable}
\caption{Baseline Parameters in Yield Modeling and Simulation}
\label{tab:model_parameters}
\begin{tabular}{@{}ll@{}}
\toprule
\textbf{Design Parameters} & \textbf{Value} \\
\midrule
Pad pitch~\cite{Zhao2024} & \SI{1}{\micro\meter} \\
Bottom/Top pad size~\cite{Zhao2024} & \SI{0.5}{\micro\meter}, \SI{0.3}{\micro\meter} \\
Die size~\cite{Dylan2024} & \SI{10}{\milli\meter}$\times$\SI{10}{\milli\meter} \\
Wafer size & \SI{300}{\milli\meter} \\
\midrule
\textbf{Process Parameters} & \textbf{Value} \\
\midrule
Random misalignment~\cite{Ryan2025} & \SI{0}{\nano\meter} (\SI{20}{\nano\meter})\tnote{*} \\
System x, y translation~\cite{Ryan2025} & \SI{0}{\nano\meter} (\SI{20}{\nano\meter})\tnote{*} \\
System rotation~\cite{Ryan2025} & \SI{0.05}{\micro\radian} (\SI{0.01}{\micro\radian})\tnote{*} \\
System magnification~\cite{Ryan2025} & \SI{0.05}{ppm} (\SI{0.01}{ppm})\tnote{*} \\
Particle defect density~\cite{Dylan2024} & \SI{0.1}{\centi\meter^{-2}} \\
Minimum particle thickness~\cite{Dylan2024} & \SI{0.1}{\micro\meter} \\
Shaping factor $z$ in Eq.~\ref{eq:defect_thickness_distribution}~\cite{Stapper1984, Bruls1992} & 3 \\
Top/Bottom pad recess~\cite{Kim2020, Ji2020} & \SI{10}{\nano\meter} (\SI{1}{\nano\meter})\tnote{*} \\
Roughness $\sigma_z$~\cite{Dubey2024,Dubey2023} & \SI{1}{\nano\meter} \\
Adhesion energy (SiO$_2$--SiO$_2$)~\cite{Chidambaram2020, Fujii2023} & \SI{1.2}{\joule\per\meter\squared} \\
Young's modulus (SiO$_2$)~\cite{Ji2020,Zhao2024} & \SI{73}{\giga\pascal} \\
Dielectric thickness~\cite{Chidambaram2021} & \SI{1.5}{\micro\meter} \\
\midrule
\textbf{Model Parameters} & \textbf{Value} \\
\midrule
Contact area constraint $k_{ca}$ in Eq.~\ref{eq:circ_delta}~\cite{Ryan2025} & 0.5 \\
Critical distance constraint $k_{cd}$ in Eq.~\ref{eq:circ_delta}~\cite{Ryan2025} & 0.5 \\
$k_{mag}$ in Eq.~\ref{eq:bow_model}~\cite{Kang2024} & \SI{0.09}{\meter^{-1}} \\
$k_{peel}$ in Eq.~\ref{eq:linear_peeling_stress_model}~\cite{Ji2020} & \SI{6.55e15}{\newton\cdot\meter^{-3}} \\
$h_0$ in Eq. \ref{eq:linear_peeling_stress_model}~\cite{Ji2020} & \SI{75}{\nano\meter} \\
$k_r$ in Eq. \ref{eq:void_size_model}~\cite{Nagano2022} & \SI{1.8e-4}{\micro\meter^{-1/2}}\\
$k_{r0}$ in Eq. \ref{eq:void_size_model}~\cite{Nagano2022} & \SI{230}{\micro\meter^{1/2}} \\
$k_l$ in Eq. \ref{eq:void_tail_length_model}~\cite{Nagano2022} & \SI{6.2e-2}{\micro\meter^{-1/2}} \\
$k_n$ in Eq. \ref{eq:num_void_in_void_tail_model}~\cite{Nagano2022} & \SI{9e-5}{\micro\meter^{-3/2}} \\
$k_S$ in Eq. \ref{eq:void_tail_area_model}~\cite{Nagano2022} & \SI{2.7}{\micro\meter^{1/2}} \\
W2W gridding resolution$^\dag$ & \SI{400}{\micro\meter}$\times$\SI{400}{\micro\meter} \\
D2W gridding resolution$^\dag$ & \SI{100}{\micro\meter}$\times$\SI{100}{\micro\meter} \\
\bottomrule
\end{tabular}
\begin{tablenotes}
\item[*] Mean (Standard Deviation)
\item[\dag] Gridding resolution is used exclusively in the analytical modeling and does not affect the simulation results.
\end{tablenotes}
\end{threeparttable}
\end{table}
\subsection{Overlay Check}
Regarding overlay errors, we assume the random misalignment and three distortion components (translation, rotation, and magnification errors) follow respective normal distributions. 
In each input parameter set, multiple combinations of those values are sampled from the distributions and are used to calculate the overall misalignment. 
In W2W simulations, the parameters are sampled per wafer, whereas in D2W simulations, they are sampled per die.
The Cu connection fails if the overall misalignment exceeds the maximum allowed overlay error $\delta$.

\subsection{Defect Check}
We initially assign thicknesses to the particles by randomly sampling the thickness distribution $D(t)$ given in Eq.~\ref{eq:defect_thickness_distribution}.
Although the number of particle defects may vary between individual wafers/dies, their occurrence across the wafer/die population follows the particle defect density $D_t$.
Then, the particles are randomly and uniformly located across the wafer/die, and the void tails are generated based on the linear fitting model from \cite{Nagano2022}, simulating the bond wave propagation (Fig. \ref{fig:void_visualizer}).
The main void size and the void tail length are given by Eq. \ref{eq:void_size_model} and Eq. \ref{eq:void_tail_length_model}, respectively. 
To accurately simulate the defect morphology, it is necessary to account for both the number and size distribution of voids in the void tail.
The number of voids in the void tail is given by
\begin{equation}
    n = k_nLt^{1/2}
    \label{eq:num_void_in_void_tail_model}
\end{equation}
The total area of the void tail is given by
\begin{equation}
    S=k_SLt^{1/2}
    \label{eq:void_tail_area_model}
\end{equation}
Within the void tail, the size of each void decreases linearly as its position shifts farther toward the wafer edge. The Cu connection fails if there is any void overlapping with the top pad.

\begin{figure}[t]
    \centering
    \begin{subfigure}[b]{0.49\linewidth}
    \centering
    \includegraphics[width=\linewidth]{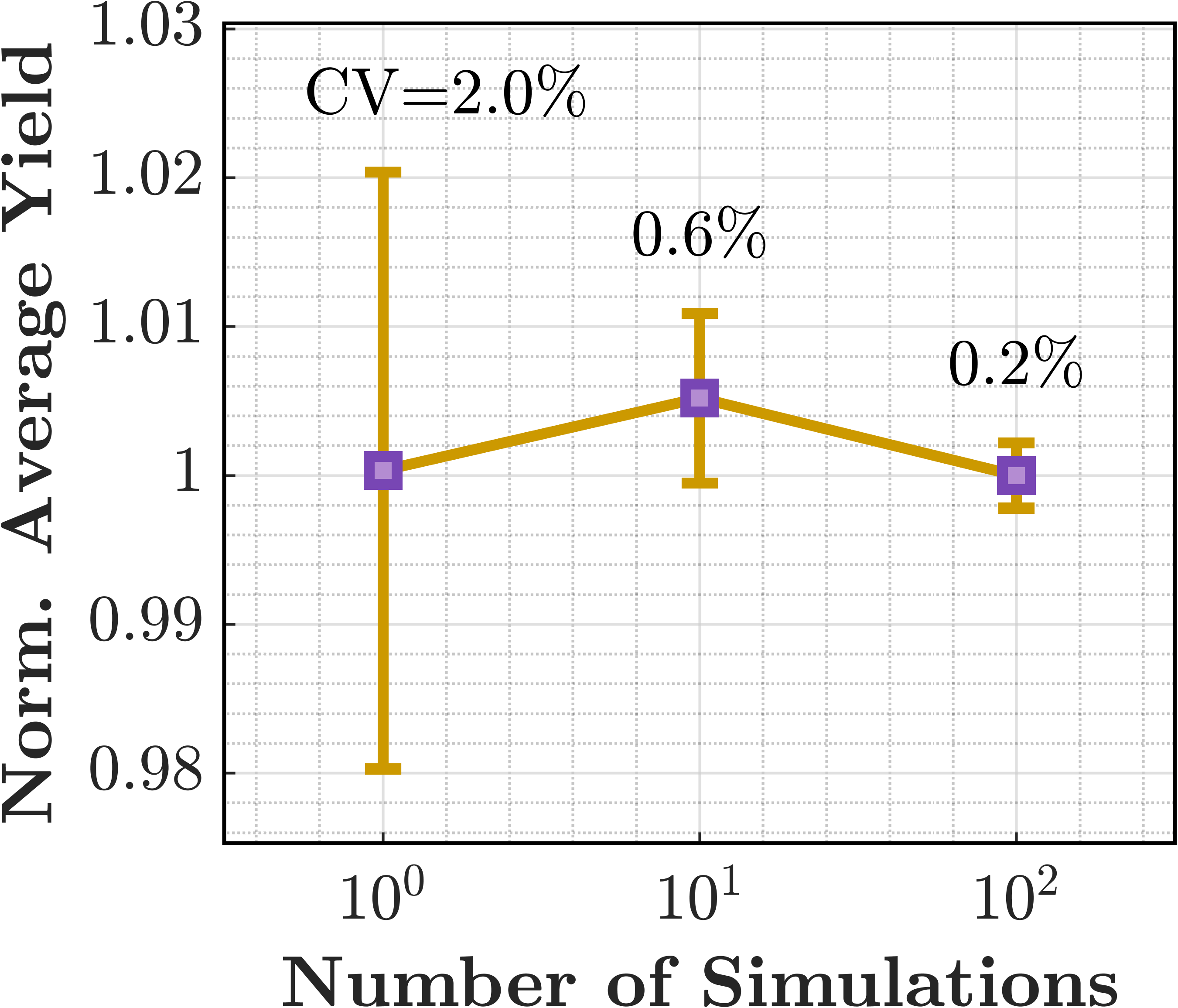}
    \caption{W2W HB.}
    \label{fig:w2w_normalized_simulation_cv}
    \end{subfigure}
    \hfill
    \begin{subfigure}[b]{0.49\linewidth}
        \centering
        \includegraphics[width=\linewidth]{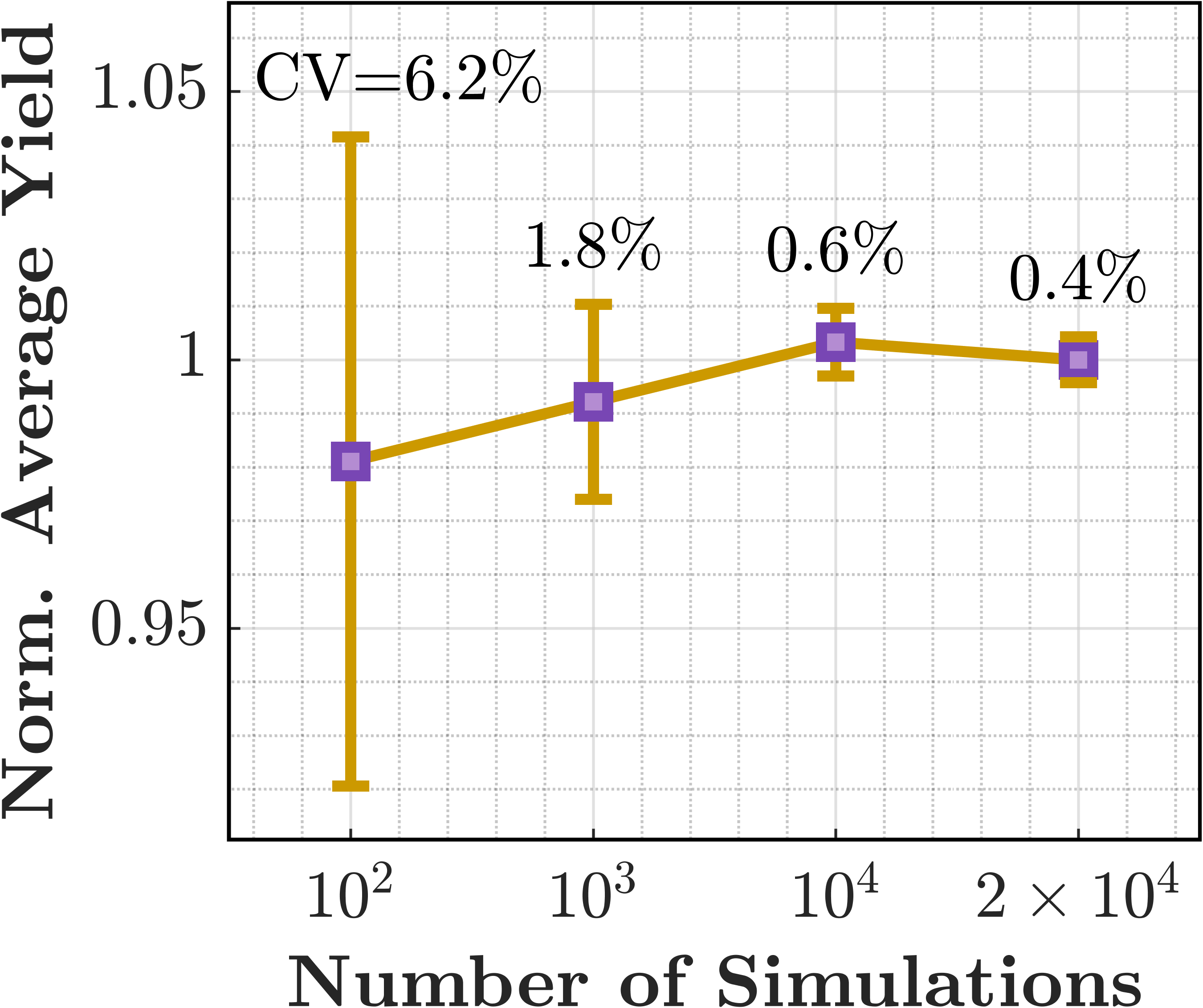}
        \caption{D2W HB.}
        \label{fig:d2w_normalized_simulation_cv}
    \end{subfigure}
    \caption{Simulation effort analysis for reliable yield results.}
    \label{fig:sim_efforts}
\end{figure}
\begin{figure}[t]
    \centering
    \begin{subfigure}[b]{0.49\linewidth}
        \centering
        \includegraphics[width=\linewidth]{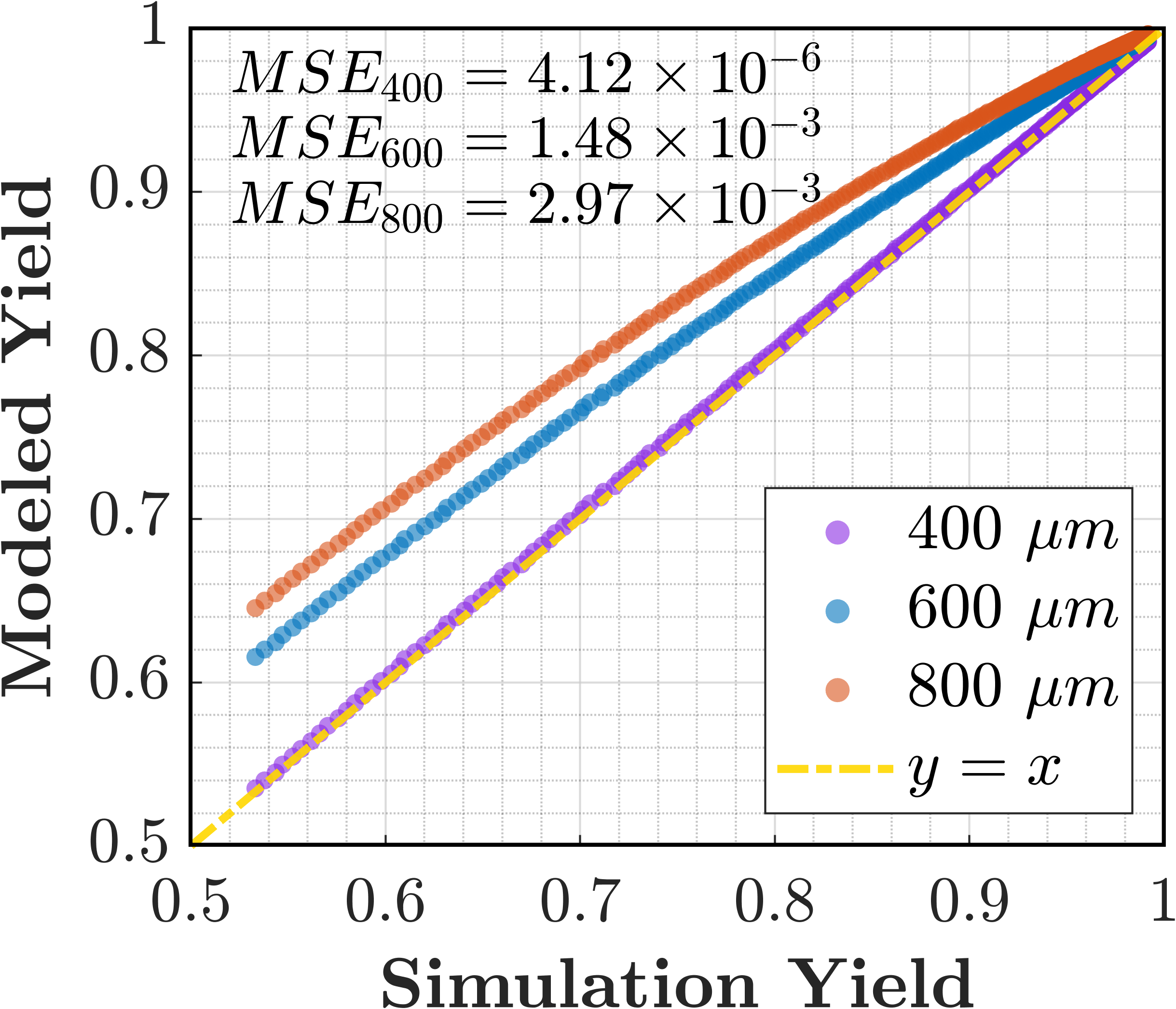}
        \caption{Defect model of W2W HB.}
        \label{fig:w2w_pad_block_dim_impact_on_Ydf}
    \end{subfigure}
    \hfill
    \begin{subfigure}[b]{0.49\linewidth}
        \centering
        \includegraphics[width=\linewidth]{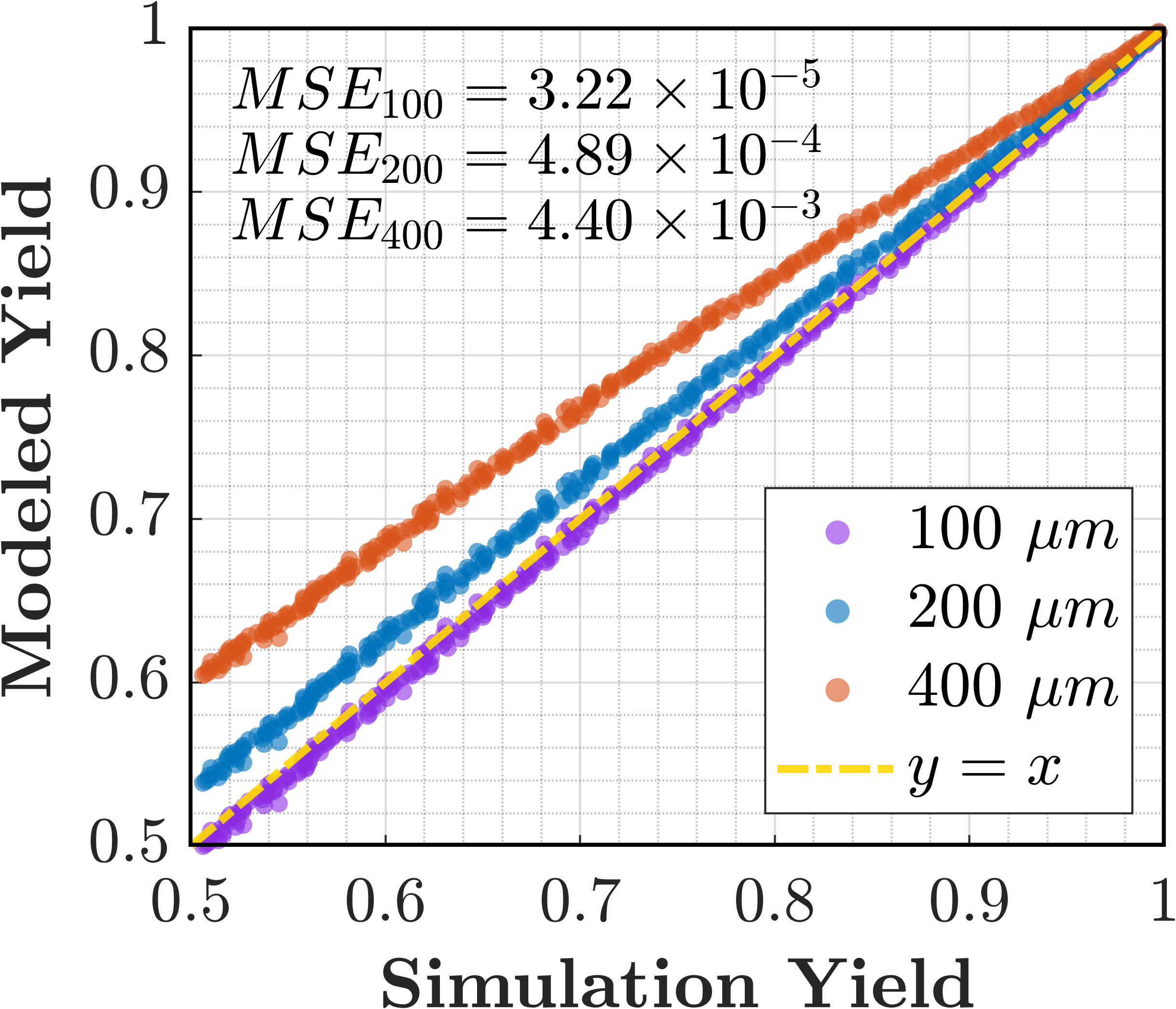}
        \caption{Defect model of D2W HB.}
        \label{fig:d2w_pad_block_dim_impact_on_Ydf}
    \end{subfigure}
    \caption{Correlation between 3 gridding-resolution-based defect yield modeling results and 1 reference simulation result. The gridding resolution in modeling does not affect the simulation results.}
    \label{fig:pad_block_dim_impact_on_Ydf}
\end{figure}

\begin{figure}[t]
    \centering
    \begin{subfigure}[b]{0.49\linewidth}
        \centering
        \includegraphics[width=\linewidth]{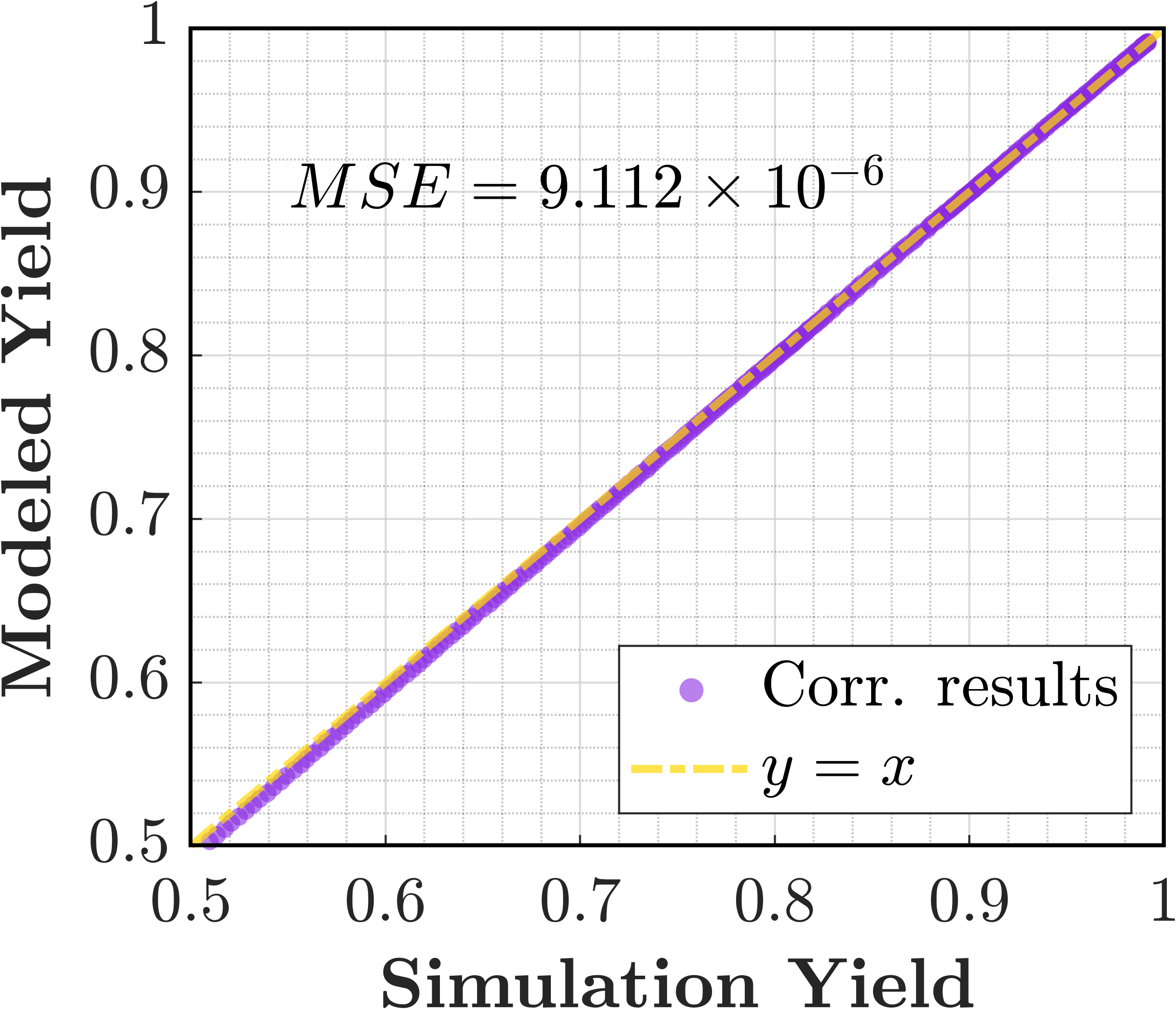}
        \caption{Yield of W2W HB.}
        \label{fig:w2w_assembly_correlation}
    \end{subfigure}
    \hfill
    \begin{subfigure}[b]{0.49\linewidth}
        \centering
        \includegraphics[width=\linewidth]{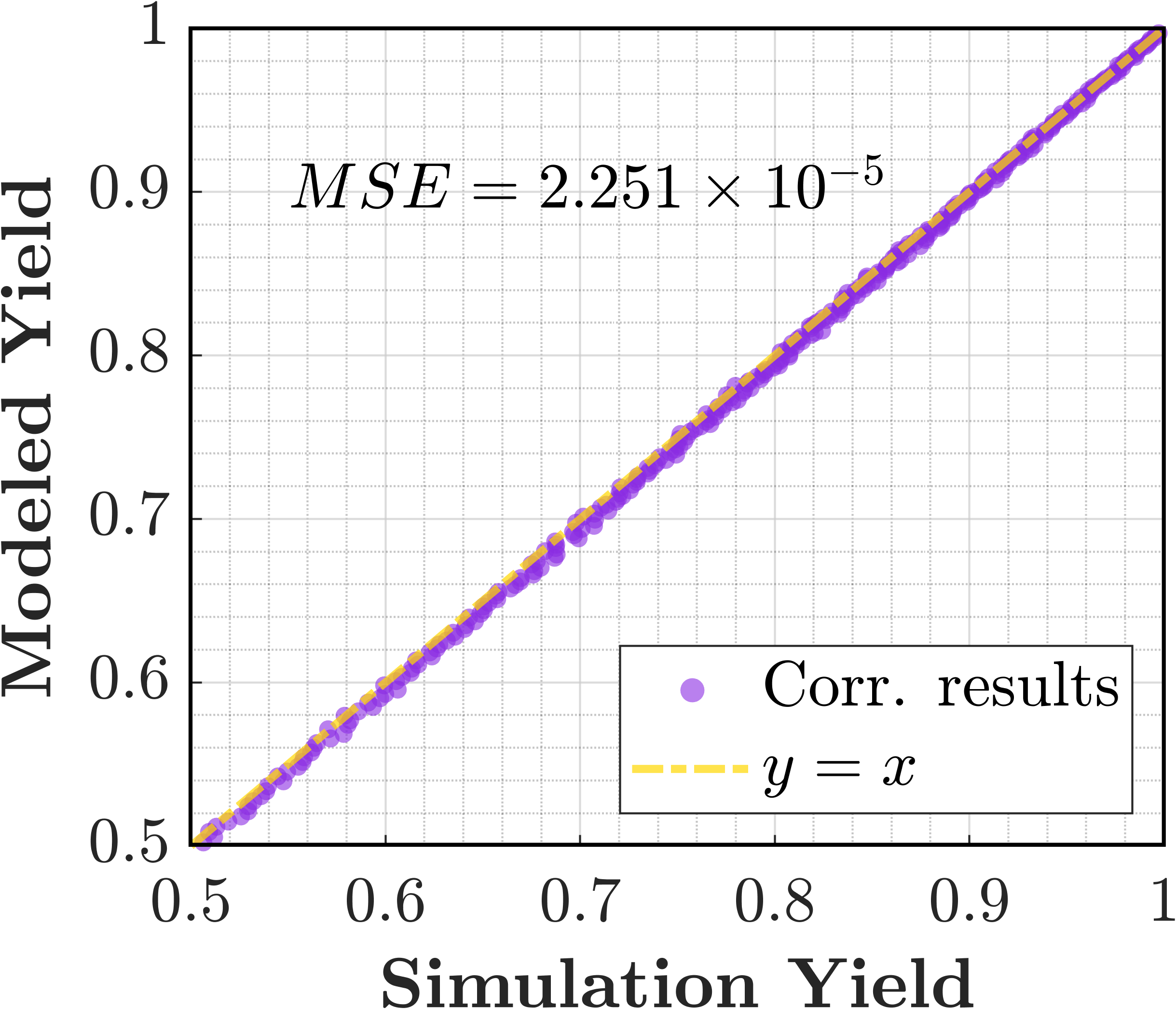}
        \caption{Yield of D2W HB.}
        \label{fig:d2w_assembly_correlation}
    \end{subfigure}
    \caption{Correlation of bonding yield with the simulation.}
    \label{fig:assembly_correlation}
\end{figure}

\begin{table}[t]
\small
\centering
\caption{Runtime of the defect yield across different gridding resolutions (GR) for W2W and D2W HB}
\label{tab:gridding_resolution_vs_runtime}
\begin{tabular}{cc|cc}
\hline
\multicolumn{2}{c|}{W2W} & \multicolumn{2}{c}{D2W} \\
\cline{1-4}
GR (\SI{}{\micro\meter})& $Y_{df}$ runtime (s) & GR (\SI{}{\micro\meter}) & $Y_{df}$ runtime (s) \\
\hline
800 & 0.10  & 400 & 0.25 \\
600 & 0.11 & 200 & 1.17 \\
400 & 0.14 & 100 & 7.76 \\
200 & 0.38 & 50 & 65.97 \\
\hline
\end{tabular}
\end{table}
\subsection{Cu Recess Check}
The Cu recess values of the top pad and bottom pad are sampled from their respective normal distributions.
The peeling stress $\sigma_{peel}$ during PBA and the gap between the two Cu pads after PBA are calculated.
The Cu connection fails if: (1) the peeling stress is higher than the tolerance value, or (2) the gap still exists after PBA.
\begin{figure*}[t]
    \centering    
    \includegraphics[width=\textwidth]{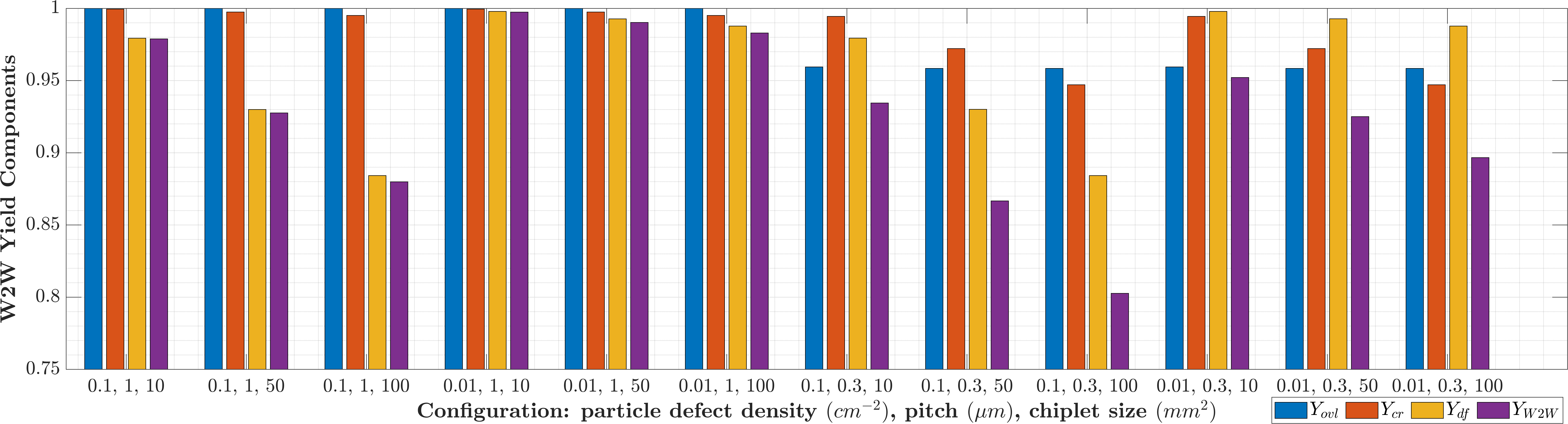}
    \caption{W2W case studies for various configurations.}
    \label{fig:w2w_bar}
\end{figure*}
\begin{figure*}[t]
    \centering    
    \includegraphics[width=\textwidth]{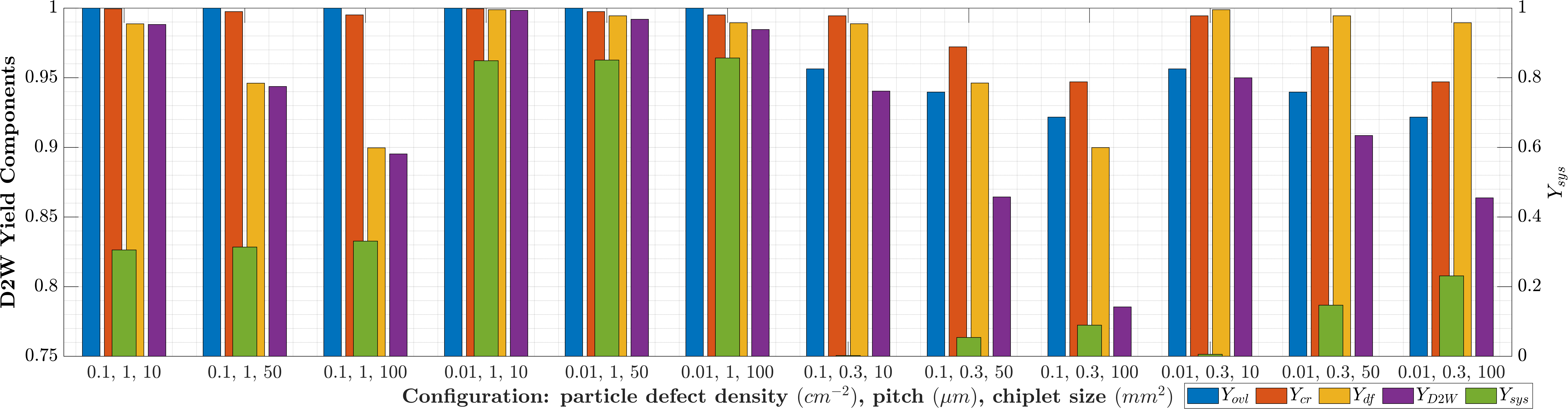}
    \caption{D2W case studies for various configurations.}
    \label{fig:d2w_bar}
\end{figure*}
To summarize, a Cu connection survives only if it passes the \textit{Overlay Check}, \textit{Defect Check}, and \textit{Cu Recess Check}.
A die is considered to have survived if (1) all critical pads remain functional, and (2) in each group of redundant pads, at least one pad survives.
The simulation results are closer to the actual conditions with less approximation compared to the model. 
However, to achieve accurate yield predictions across all failure mechanisms using baseline inputs, repeated simulations are necessary to obtain a statistically reliable mean yield.
As illustrated in Fig. \ref{fig:sim_efforts}, we perform multiple simulations for a single input parameter set and compute the average yield. This process is repeated 10 times to observe the variability of the results.
The coefficient of variation (CV), defined as the ratio of the standard deviation to the mean, is annotated for each simulation count to quantify the relative dispersion.
We determine the minimum number of simulations required to reduce the CV below 1\% for both the W2W and D2W cases, and use this simulation setting to validate the YAP+ model.
It requires 10 wafer samples (10,000 die samples) for W2W (D2W) HB simulation, taking 0.9 hours (2.0 hours) on a single CPU (AMD Ryzen 9 8945HS). 
Additionally, since dilation accounts for a significant portion of the model’s runtime, our model validation examines the impact of a key factor in this process, the gridding resolution, on model accuracy.
Gridding resolution plays a crucial role in determining the accuracy of defect yield estimation.
As shown in Fig. \ref{fig:pad_block_dim_impact_on_Ydf}, the acceptable gridding resolutions for W2W and D2W hybrid bonding are \SI{400}{\micro\meter}$\times$\SI{400}{\micro\meter} and \SI{100}{\micro\meter}$\times$\SI{100}{\micro\meter}, respectively.
These settings are used as the baseline configurations in the subsequent experiments.
A coarser grid reduces model accuracy by inadequately representing the pad layout and defect geometry, which compromises the resolution of the critical area, resulting in inaccurate defect yield estimation.
Conversely, using an excessively fine grid offers only marginal accuracy improvement while significantly increasing the dilation runtime, as shown in Table \ref{tab:gridding_resolution_vs_runtime}.
By using the baseline in Table \ref{tab:model_parameters}, the yield model achieves virtually identical accuracy compared with simulation results, as shown in Fig. \ref{fig:assembly_correlation} in 3.7 s (10.2 s) for W2W (D2W) HB, offering over \textit{870x (720x) runtime improvement}.

If the computational overhead of the dilation-based critical area calculation in the defect model is amortized through the use of a look-up table, the runtime can be further reduced to 2.9 s (2.3 s) and can achieve over \textit{1,100x (3,200x) runtime improvement} for W2W (D2W) HB.
The high modeling efficiency enables the usage in yield optimization and pathfinding optimization loops.

\label{3_methodology}

\section{Experimental Results}
After validating the model with simulation results, we use YAP+ to conduct case studies that demonstrate the impact of various process and design factors on bonding yield, indicating its strengths in system-technology co-optimization. 
We vary particle defect density (\SI{0.01}{\per\square\centi\meter}, \SI{0.1}{\per\square\centi\meter}), pitch (\SI{0.3}{\micro\meter}, \SI{1}{\micro\meter}), and chiplet sizes (\SI{10}{\milli\meter\squared}, \SI{50}{\milli\meter\squared}, \SI{100}{\milli\meter\squared}) in the modeling. 
A pad layout composed exclusively of critical pads is adopted in this experiment.
The yield breakdown and the overall bonding yield are reported in Fig.\ref{fig:w2w_bar} (W2W setup) and Fig.\ref{fig:d2w_bar} (D2W setup).

\subsection{The Impact of Particle Defect Density}
The HB process requires the strict removal of particles at the bonding interface. 
Fig. \ref{fig:w2w_bar}, \ref{fig:d2w_bar} show that under a relaxed bonding pitch (\SI{1}{\micro\meter}), bonding yield is notably affected by defect-related failures.
W2W HB exhibits higher sensitivity to particle contamination due to void tail formation during bond wave propagation, resulting in a larger critical area per die.
The results indicate that a 10x improvement in defect density (ISO 2) enables near-perfect (\SI{\sim 99}{\percent}) defect yield for both W2W and D2W across all chiplet sizes. 

\subsection{Impact of Bonding Pitch}
In this case study, the bottom pad size is set to half the corresponding pitch.
As shown in Fig.\ref{fig:w2w_bar} and Fig.\ref{fig:d2w_bar}, reducing the pitch from \SI{1}{\micro\meter} to \SI{0.3}{\micro\meter} leads to a noticeable drop in yield across various chiplet sizes, with the effect more pronounced in D2W HB.
Smaller pitches increase sensitivity to Cu pillar misalignment, demanding tighter overlay control.
Currently, both W2W and D2W HB technologies are capable of achieving \SI{50}{\nano\meter} overlay accuracy~\cite{Ryan2025, Sano2025}.
However, in D2W HB, the overlay yield ($Y_{ovl}$) at a \SI{0.3}{\micro\meter} pitch gradually degrades as the chiplet size increases. 
In contrast, W2W HB maintains a relatively stable overlay yield across all chiplet sizes.
This distinction, given the comparable alignment accuracy, arises from how overlay error affects yield in these two bonding schemes. 
In D2W HB, if the alignment error at the chiplet edge exceeds the failure threshold, the die is discarded.
However, in W2W HB, chiplets located near the wafer center are more likely to survive even when edge alignment reaches the failure limit, thus resulting in higher $Y_{ovl}$.

Reducing the bonding pitch significantly increases the number of I/O pads, heightening sensitivity to Cu recess variations. 
As shown, the yield loss in W2W HB at smaller pitches is mainly due to reduced $Y_{ovl}$ for smaller chiplets and reduced $Y_{cr}$ for larger chiplets.
A practical strategy to mitigate the impact of increasing I/O pad counts is to introduce redundancy to critical pads.
The defect yield, on the other hand, remains largely unaffected, as the void sizes exceed the pitch, keeping the critical area roughly constant.

\subsection{Analyzing Yield Limiters with Varying Chiplet Sizes}
Bonding yield decreases with increasing chiplet size for both D2W and W2W bonding, primarily due to greater Cu recess variation (from more I/O pads per die) and heightened defect sensitivity.

D2W hybrid bonding can be applied in 2.5D integration to assemble large chiplet systems. 
As such, evaluating yield based solely on a single chiplet can be misleading.
Although full system-level yield modeling is beyond the scope of this work, we include a system yield ($Y_{sys}$, shown in Fig. \ref{fig:d2w_bar}), calculated as $Y_{D2W}^{\#chiplets}$, assuming no chiplet redundancy and a nominal system size of \SI{1000}{\milli\meter\squared}.
Note that $Y_{sys}$ is plotted against the right vertical axis.
This approach reflects the cumulative probability of successful bonding across all chiplets in the system. 
Increasing chiplet size reduces the total number of chiplets required, helping to mitigate the compounding effect of $Y_{D2W}$ degradation on the $Y_{sys}$. 
Interestingly, even though $Y_{D2W}$ decreases with increasing chiplet size, the system yield $Y_{sys}$ remains slightly higher.
\footnote{Note that this does not account for any worsened yield of a larger chiplet. A more complete system yield model can be found, for example, in \cite{Graening2023}, albeit with an oversimplified bonding yield model.}
Overall, building large chiplet-based systems using 2.5D integration necessitates tighter process control, particularly in overlay alignment and Cu recess variations.
\begin{figure}[t]
    \centering
    \begin{subfigure}[b]{\linewidth}
        \centering
        \includegraphics[width=\linewidth]{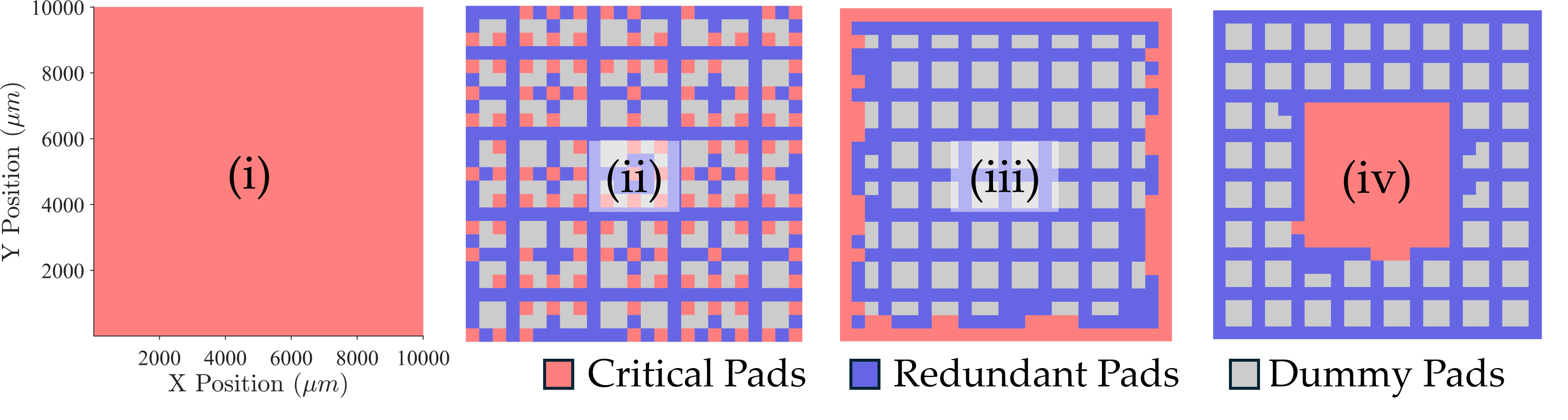}
        \caption{I/O pad layouts: (i) \textit{Full}; (ii) \textit{Sparse}; (iii) \textit{Peripheral}; (iv) \textit{Centralized}.}
        \label{fig:multi_pad_layout_pattern}
    \end{subfigure}
    \begin{subfigure}[b]{\linewidth}
        \centering
        \includegraphics[width=\linewidth]{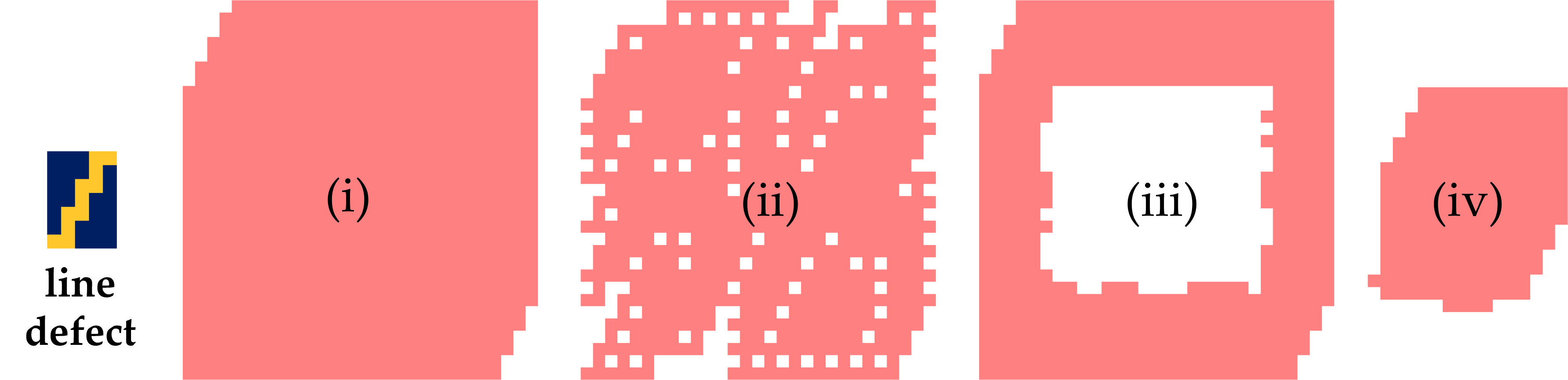}
        \caption{Comparison of critical areas in W2W HB across different pad layouts, evaluated with a void tail defect of \SI{3}{\milli\meter} in length and oriented at \SI{30}{\degree}.}
        \label{fig:pad_layout_critical_area}
    \end{subfigure}
    \begin{subfigure}[b]{\linewidth}
        \centering
        \includegraphics[width=\linewidth]{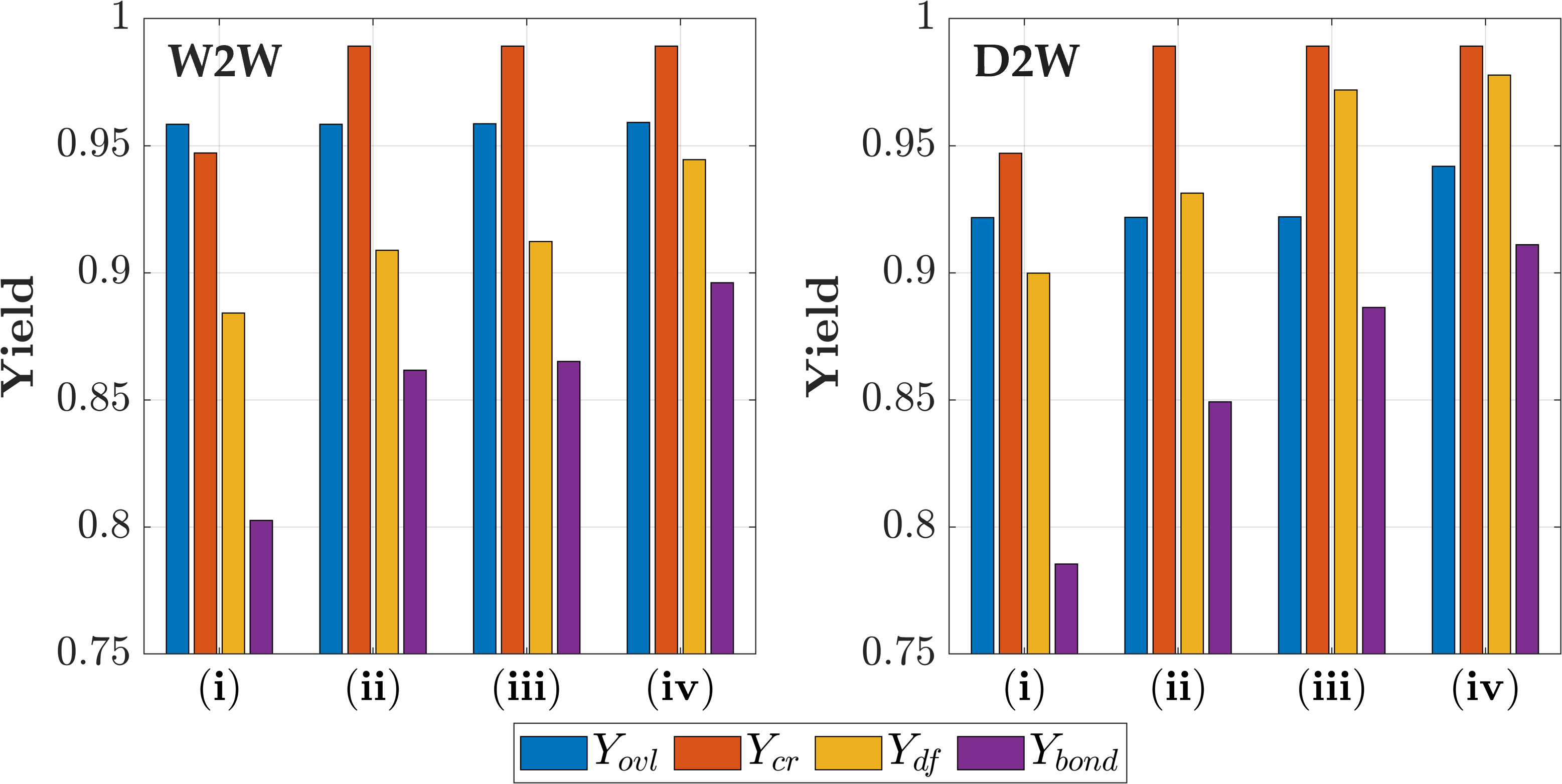}
        \caption{Yield breakdown of W2W and D2W HB across I/O pad layouts.}
        \label{fig:yield_breakdown_pad_layouts}
    \end{subfigure}
    \caption{Impact of I/O pad layouts on W2W and D2W bonding yield at \SI{0.3}{\micro\meter} pitch.}
    \label{fig:impact_of_pad_layout}
\end{figure}
\subsection{The Impact of the I/O Pad Layout}
The layout of I/O pads significantly affects bonding yield in both W2W and D2W HB, as the failure of different types of I/O pads has different contribution mechanisms to the yield loss.
To investigate this, four distinct I/O pad spatial distribution patterns are studied:
\begin{enumerate}[label=(\roman*)]
    \item \textit{Full}: all pads are critical, evenly distributed across the die area;
    \item \textit{Sparse}: critical pads are spread sparsely with additional redundant pads and dummy pads; 
    \item \textit{Peripheral}: critical pads are arranged around the edges with redundant and dummy pads inside. 
    \item \textit{Centralized}: critical pads are concentrated at the center, surrounded by redundant and dummy pads.
\end{enumerate}
Fig. \ref{fig:multi_pad_layout_pattern} visualizes these four patterns.
These I/O pad layouts are configured at \SI{0.3}{\micro\meter} pitch to highlight their impact on $Y_{ovl}$ and $Y_{cr}$, as yield loss due to overlay errors and Cu recess variations is more pronounced at finer pitches.
\textit{Full} layout consists of 100\% critical pads, while the remaining layouts each contain 20\% critical pads, 50\% redundant pads, and 30\% dummy pads.
\textit{Peripheral} layout is widely adopted in 2.5D integration, where die-to-die interconnections require placing I/O pads at the die periphery to simplify routing.
\textit{Centralized} layout partially resembles that of 3D-stacked memory systems like HBM, where a horizontally distributed central TSV region delivers signals and power vertically and is surrounded by multiple memory channels \cite{Park2022}.
For D2W HB, a pad block size of \SI{100}{\micro\meter}$\times$\SI{100}{\micro\meter} is adopted consistently across all layouts. 

Fig. \ref{fig:pad_layout_critical_area} depicts the critical areas of these pad layouts, evaluated with a void tail defect in W2W HB, under the same scale for direct comparison.
Fig. \ref{fig:yield_breakdown_pad_layouts} shows the yield breakdown across four layouts.
Overall, compared to \textit{Full} layout, the other three have higher Cu recess yield $Y_{cr}$ and defect yield $Y_{df}$.
This improvement arises from the reduced number of critical pads, which lowers the cumulative impact of individual pad failures due to Cu recess variations and slightly decreases the critical area vulnerable to particle-induced defects.
Within three layouts with less critical pads, \textit{Sparse} layout results in relatively low $Y_{df}$ improvement, as its critical pads are more spatially isolated. 
As the pattern (ii) shown in Fig. \ref{fig:pad_layout_critical_area}, this reduces the overlap among the critical areas of adjacent pad blocks, leading to a larger cumulative critical area. 
In contrast, \textit{Peripheral} and \textit{Centralized} layouts feature more clustered critical pads, where overlapping of their critical areas effectively `offsets' the total exposed critical region, thus enhancing defect tolerance.

In the W2W case, \textit{Centralized} layout achieves over 3\% improvement in $Y_{df}$ compared to \textit{Peripheral} layout. Meanwhile, for D2W, the two layouts exhibit similar $Y_{df}$ values, with a difference of less than 0.6\%.
This trend can be attributed to the differing defect morphologies in W2W and D2W bonding. 
In the W2W case, the dominant defect type is the void tail, which has a relatively large spatial footprint. 
As a result, the extended dimension of void tails occupies a significant portion of the critical area in the internal cavity in \textit{Peripheral} layout, as the pattern (iii) in Fig. \ref{fig:pad_layout_critical_area}, limiting the effectiveness of critical area reduction. 
In contrast, \textit{Centralized} layout benefits from the tight clustering of critical pad blocks, which maximizes the overlap among their associated critical areas. 
This overlap acts as an effective compression or offset mechanism, substantially reducing the net defect-sensitive region and leading to higher $Y_{df}$.
In the D2W case, the dominant defect is the main void, which is more localized and has a smaller spatial dimension. 
Under this condition, both \textit{Peripheral} and \textit{Centralized} layouts achieve comparable levels of critical area overlap, resulting in $Y_{df}$ values.
However, to mitigate the impact of debris contamination from edge chipping during die dicing, a \textit{Centralized} layout is the recommended choice \cite{Xie2025}. 
Besides offering a higher $Y_{df}$, the \textit{Centralized} layout also contributes to improved $Y_{ovl}$ in D2W hybrid bonding, particularly under ultra-fine pitch regimes, by mitigating maximum radial overlay errors such as those caused by rotation and magnification at the periphery of the critical pad array.

In summary, while any layout with fewer critical pads generally improves yield over a \textit{Full} layout, the most effective strategy for boosting defect yield is to use a compact, clustered distribution of critical pads. 
This approach is particularly advantageous in processes characterized by high defect densities and large defects.

\begin{figure}[t]
    \centering
    \begin{subfigure}[b]{0.49\linewidth}
    \centering
    \includegraphics[width=\linewidth]{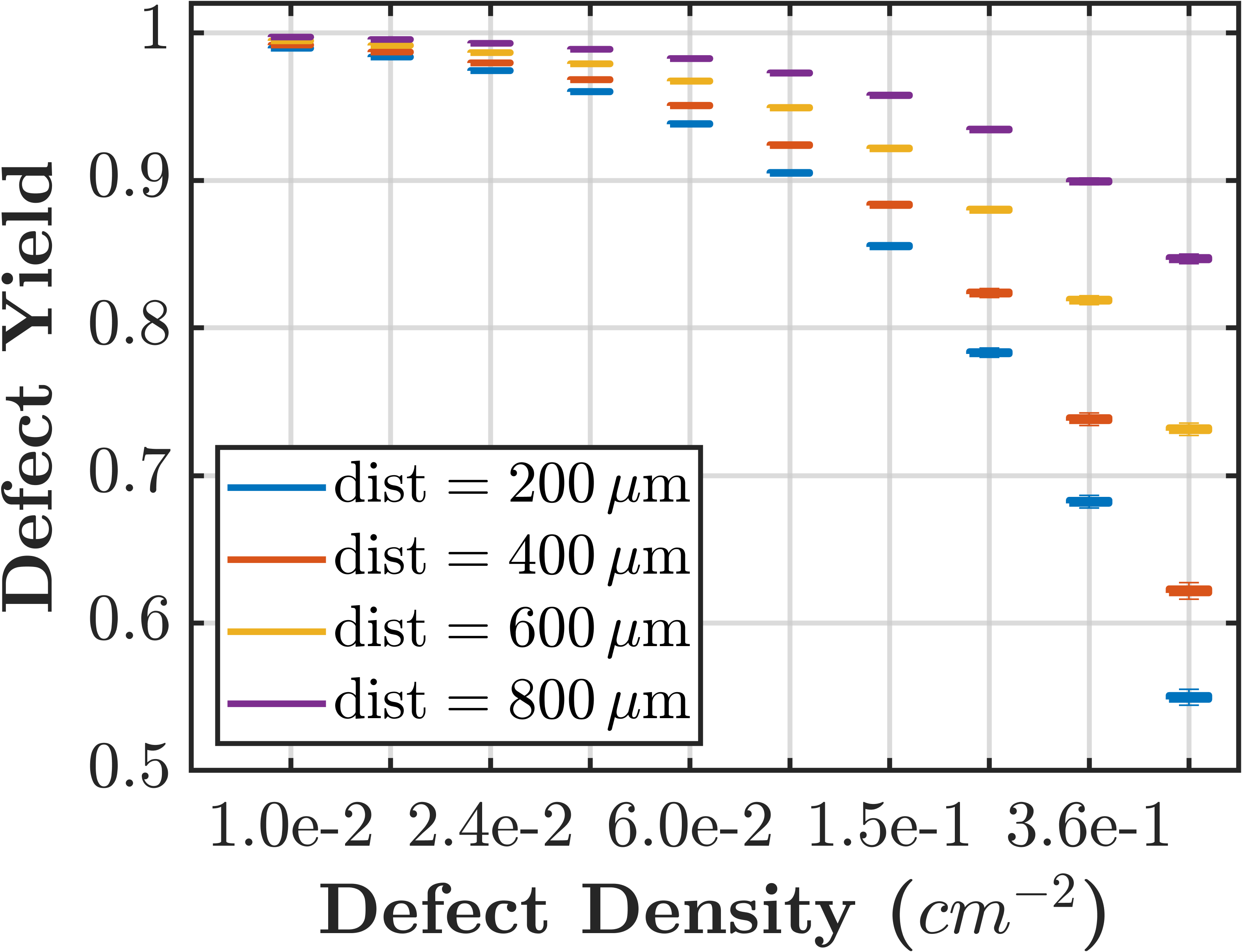}
    \caption{Die size: \SI{10}{\milli\meter}$\times$\SI{10}{\milli\meter}.}
    \label{fig:replica_dist_10x10_1000pts}
    \end{subfigure}
    \hfill
    \begin{subfigure}[b]{0.49\linewidth}
        \centering
        \includegraphics[width=\linewidth]{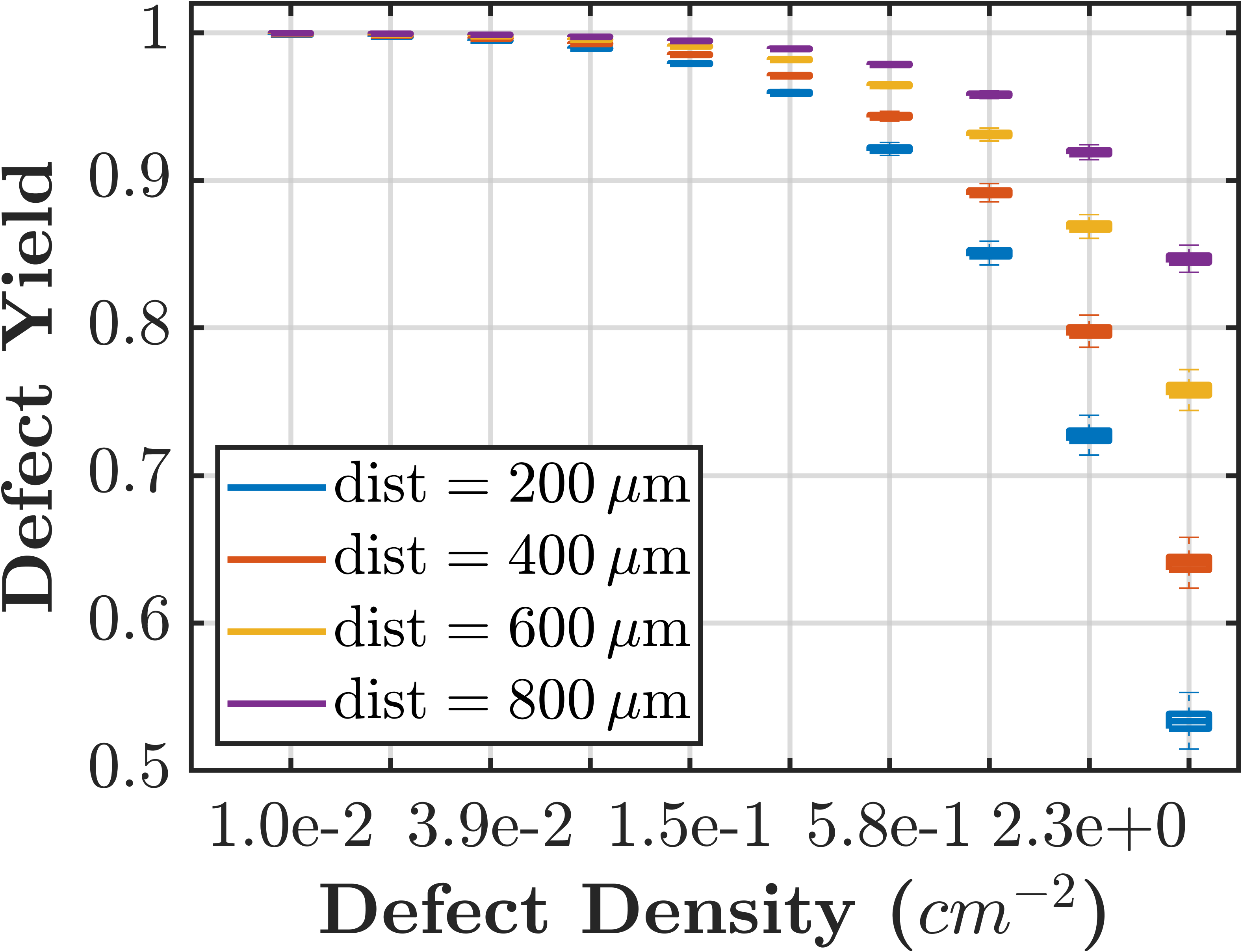}
        \caption{Die size: \SI{3.2}{\milli\meter}$\times$\SI{3.2}{\milli\meter}.}
        \label{fig:replica_dist_3d2x3d2_1000pts}
    \end{subfigure}
    \caption{Impact of the main-replica spacing on the defect yield of W2W HB across 1,000 randomly generated pad layouts.}
    \label{fig:replica_location}
\end{figure}

\subsection{The Impact of the Redundant Replicas}
In many IC chips, identical blocks of circuits are often replicated to enhance yield \cite{Koren1998}. 
For fine-pitch, large-scale designs with a high number of I/O pads, numerous sparsely distributed dummy resources can be allocated for redundancy. This flexibility allows designers to adopt either shared or dedicated redundancy schemes.
Adding redundancy provides a straightforward improvement in Cu recess yield $Y_{cr}$, with gains directly tied to the number of redundant pairs.
However, the impact on defect-related yield, $Y_{df}$, depends not only on the redundancy strategy but also on the main-replica spacing, the physical separation between the primary pad and its replica, and the size of defects.
In HB process, defect sizes, whether from void tails or main voids, are typically more than 100 times the pad pitch. 
In such cases, shared redundancy is generally ineffective because the small spacing between pads in the same group makes them prone to simultaneous failure.
To achieve meaningful yield gains, dedicated redundancy with sufficiently large spacing between paired pads is necessary. 
The influence of main-replica spacing on $Y_{df}$ is explored below.

In this experiment, we specify a pad block size of \SI{200}{\micro\meter}$\times$\SI{200}{\micro\meter} for both W2W and D2W HB.
To eliminate the influence of critical pads, the die is composed entirely of 100\% redundant pads.
Among the redundant blocks, half are designated as main pad blocks and the other half as their corresponding replica pad blocks, where the replica pads are located.
The redundancy scheme adopts a 1:1 mapping between main pads and replica pads.
In the pad block assignment, the main pad blocks are randomly distributed across the die.
For each main pad block, the algorithm searches for candidate blocks located at a specified Euclidean distance and assigns one as the replica pad block. 
This process continues until all replicas are paired.

\begin{figure}[t]
    \centering
    \includegraphics[width=\linewidth]{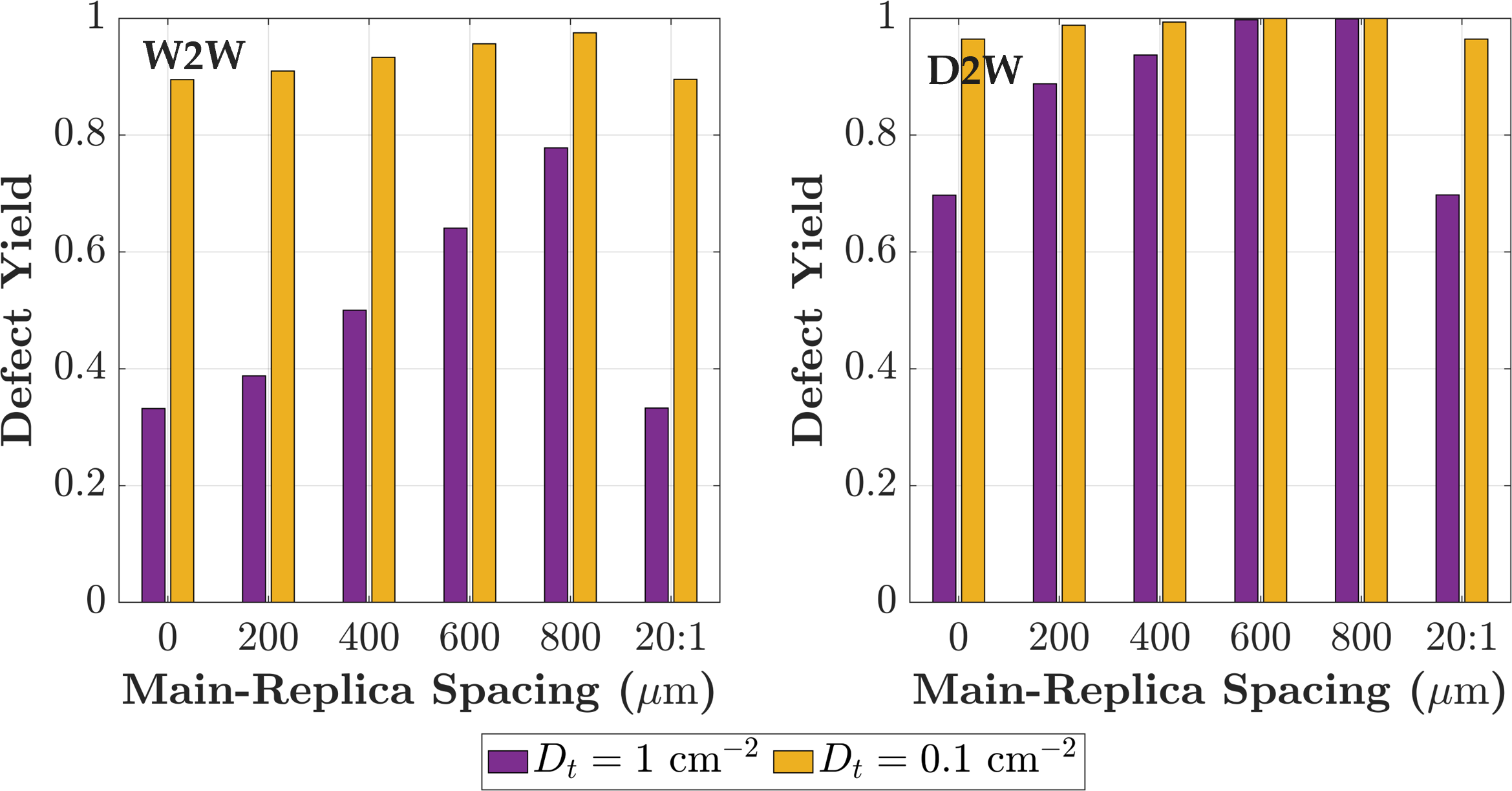}
    \caption{Defect yield of W2W and D2W HB under various redundant replica configurations.}
    \label{fig:replica_dist_impact}
\end{figure}
We first investigate the impact of the physical locations of replicas on the defect-related yield, $Y_{df}$.
For both a \SI{10}{\milli\meter}$\times$\SI{10}{\milli\meter} die and a \SI{3.2}{\milli\meter}$\times$\SI{3.2}{\milli\meter} die, 1,000 pad layouts are randomly generated with a fixed main-replica spacing, and yield is evaluated under varying defect densities to assess the benefit of adding redundancy.
Taking W2W HB as an example, the $Y_{df}$ results shown in Fig. \ref{fig:replica_location} demonstrate that defect yields across these 1,000 layouts are tightly clustered.
This indicates that the main-replica spacing significantly impacts yield, while the exact placement of main and replica pads does not.
Furthermore, the benefit of larger spacing becomes more pronounced at higher defect densities.
Next, we examine the yield improvement achieved with different main-replica spacings.
As shown in Fig. \ref{fig:replica_dist_impact}, we report the $Y_{df}$ for both W2W and D2W HB across various redundancy configurations and two defect densities.
Here, a spacing of 0 indicates no redundancy, while “20:1” denotes a shared redundancy strategy where 20 main pads share one replica.
In W2W HB, redundancy consistently improves yield at both low and high defect densities, as this process is more sensitive to particle-induced defects compared to D2W HB.
As the main-replica spacing increases from \SI{200}{\micro\meter} to \SI{800}{\micro\meter}, the yield improvement for W2W becomes more significant.
Conversely, in D2W HB, the improvement diminishes with increasing spacing.
This behavior correlates with the characteristics of dominant defects.
In W2W HB, the primary issue is void tail defects, which can span several millimeters, as shown in Fig. \ref{fig:void_tail_length_distribution}.
Substantial yield improvement only occurs when the main-replica spacing exceeds a certain threshold (\SI{400}{\micro\meter} in this case), allowing one pad in the pair to escape the defect region.
In contrast, D2W HB is mainly affected by main void defects, which are typically smaller than \SI{200}{\micro\meter} (see Fig. \ref{fig:main_voids_distribution}).
Thus, a spacing of \SI{200}{\micro\meter} is generally sufficient to ensure redundancy effectiveness, and increasing it further brings diminishing returns.
Moreover, the results confirm that shared redundancy does not improve $Y_{df}$, likely due to the increased risk of simultaneous failures within closely packed groups.

In summary, the defect yield is primarily influenced by main-replica spacing, not the exact locations of redundant I/O pads.
A dedicated redundancy strategy with a spacing tuned to the dominant defect size in the bonding process can significantly enhance yield.
However, since longer spacings may introduce routing delays, designers must carefully balance redundancy effectiveness and performance trade-offs during the early design stages.

\label{4_experimental_results}

\section{Conclusion}
This work presents YAP+, a pad-layout-aware yield modeling framework for W2W and D2W hybrid bonding, which is proposed as an enhanced version of YAP. 
YAP+ models overlay errors, particle-induced void defects, Cu recess variations, and analyze the interaction between the failure of critical and redundant pads and the die failure.
YAP+ is validated against a physics-inspired yield simulator.
The proposed YAP+ yield model accurately predicts bonding yield across various chiplet sizes, pitches, pad layout configurations, and process parameters, and achieves a 1,000x runtime speedup over direct simulations while maintaining negligibly small mean square error. 
Case studies using YAP+ underscore critical tradeoffs between pad layouts, bonding approaches, and redundancy schemes, and reveal distinct yield-limiting mechanisms in W2W versus D2W bonding, offering concrete guidance for process control and chiplet architecture design.

Looking ahead, we aim to: (1) extend YAP+ into system-level assembly yield modeling that integrates chiplet, TSV, and interconnect yield; (2) generalize the framework to alternative bonding technologies, such as thermal-compression bonding; (3) explore yield enhancement techniques, including adaptive pad redundancy and fault-tolerant design strategies, informed by YAP+’s insights.
\label{5_conclusion}

\section*{Acknowledgement}
This work was supported in part by CHIMES, one of the seven centers in JUMP 2.0, a Semiconductor Research Corporation (SRC) program.
\label{6_acknowledgement}


\label{7_reference}

\end{document}